\documentclass[10pt,journal,compsoc]{IEEEtran}
\ifCLASSOPTIONcompsoc
  \usepackage[nocompress]{cite}
\else
  \usepackage{cite}
\fi
\ifCLASSINFOpdf
   \usepackage[pdftex]{graphicx}
   \graphicspath{{../pdf/}{../jpeg/}}
   \DeclareGraphicsExtensions{.pdf,.jpeg,.png}
\else
   \usepackage[dvips]{graphicx}
   \graphicspath{{../eps/}}
   \DeclareGraphicsExtensions{.eps}
\fi

\usepackage{microtype}                 % use micro-typography (slightly more compact, better to read)
\PassOptionsToPackage{warn}{textcomp}  % to address font issues with \textrightarrow
\usepackage{textcomp}                  % use better special symbols
\usepackage{mathptmx}                  % use matching math font
\usepackage{times}                     % we use Times as the main font
\usepackage{cite}                      % needed to automatically sort the references
\usepackage{tabu}                      % only used for the table example
\usepackage{booktabs}                  % only used for the table example

\usepackage{hyperref}

\usepackage{float}

\usepackage{comment}
\usepackage{xcolor}
\usepackage{float}
\usepackage{amsmath}
\usepackage[thinlines]{easytable}
\usepackage{bm}
\usepackage{tikz}
\usepackage{tikz}
\usepackage{xcolor}

\usepackage{fontawesome}
\usepackage{fixltx2e}
\usepackage[normalem]{ulem}

\newcommand{\toolName}{\textit{Intercept Graph}}

\begin{document}
% \title{Global Illumination for Fun and Profit}

% \title{Interactive Radial Visualization for State Difference Comparison}

% \title{A Radial Visualization Design for Interactive Comparison of State Differences}

% \title{NAME: An Enhanced Comparative Visualization for State Difference Representation}

% \title{An Augmented Radial Visualization for Data Difference Comparison}
% \title{An Augmented Radial Visualization for State Change Comparison}

% \title{Which One Performs Better? An Novel Radial Visualization for State Change Comparison}

\title{Which One Changes More? A Novel Radial Visualization for State Change Comparison}


\author{Shaolun~Ruan,
        Yong~Wang,
        and~Qiang~Guan~% <-this % stops a space
\IEEEcompsocitemizethanks{
\IEEEcompsocthanksitem S. Ruan is with the School of Computing and Information Systems, Singapore Management University.\protect\\
E-mail: slruan.2021@phdcs.smu.edu.sg.
\IEEEcompsocthanksitem Y. Wang is with the School of Computing and Information Systems, Singapore Management University.\protect\\
E-mail: yongwang@smu.edu.sg.
\IEEEcompsocthanksitem Q. Guan is with the Department of Computer Science, Kent State University.\protect\\
E-mail: qguan@kent.edu.
}% <-this % stops an unwanted space
\thanks{Manuscript received June 29, 2022; revised xx, 2022.}}


\markboth{IEEE Transactions on Visualization and Computer Graphics,~Vol.~14, No.~8, June~2022}%
{Shell \MakeLowercase{\textit{et al.}}: Bare Demo of IEEEtran.cls for Computer Society Journals}



\IEEEtitleabstractindextext{%
\begin{abstract} 
It is common to compare state changes of multiple data items and identify which data items have changed more in various applications (e.g., annual GDP growth of different countries and daily increase of new COVID-19 cases in different regions). Grouped bar charts and slope graphs can visualize both state changes and their initial and final states of multiple data items, and are thus widely used for state change comparison. But they leverage implicit bar differences or line slopes to indicate state changes, which has been proven less effective for visual comparison. Both visualizations also suffer from visual scalability issues when an increasing number of data items need to be compared. This paper fills the research gap by proposing a novel radial visualization called Intercept Graph to facilitate visual comparison of multiple state changes. It consists of inner and outer axes, and leverages the lengths of line segments intercepted by the inner axis to explicitly encode the state changes. Users can interactively adjust the inner axis to filter large changes of their interest and magnify the difference of relatively-similar state changes, enhancing its visual scalability and comparison accuracy. We extensively evaluate Intercept Graph in comparison with baseline methods through two usage scenarios, quantitative metric evaluations, and well-designed crowdsourcing user studies with 50 participants. Our results demonstrate the usefulness and effectiveness of the Intercept Graph.

% %%%% Old version by Shaolun%%%
% Data analysis often involves comparison tasks for revealing the state differences of data items.
% With an increasing scale of data, the demand to reflect the data differences is also growing.
% Visualization has been proven a powerful method to make the viewers aware of the differences.
% However, the widely-used approaches for comparison tasks, e.g., grouped bar charts and slope graphs, suffer from visual scalability issues and the ineffective encoding for differences, making the visual comparison inaccurate and non-intuitive.
% In this paper, we present \toolName, an interactive radial visualization to facilitate the comparison of state differences.
% It provides an intuitive visualization for state differences using the encoding of \textit{length} of line segments.
% Also, we propose a useful component, i.e., the inner axis, to support the smooth filtering for large differences and the accurate comparison of close differences.
% \yong{Like what I have mentioned earlier, the inner axis itself cannot be claimed as an independent contribution.}
% We demonstrate the effectiveness and usefulness of \toolName{} by usage scenarios, metric evaluation, and user studies. The results show that \toolName\ can support a better comparison of state differences over the baseline approach.
% \yong{Need to re-write it.}

\end{abstract}

}

\maketitle

\IEEEdisplaynontitleabstractindextext

% \fi
\IEEEpeerreviewmaketitle


\IEEEraisesectionheading{\section{Introduction}\label{sec:introduction}}

% \maketitle

%%% Part 1: Background: The importance of visual comparison

% \yong{Need more references to support the importance of state change comparison.}
\IEEEPARstart{S}{tate} changes refer to the variations of data between two different time stamps, entities, categories, etc., and often have practical meanings in real applications. For example, the temperature changes of fever patients before and after taking medicines can indicate the medicine effectiveness, and the gross domestic product (GDP) changes of different countries between two adjacent years show their economic growth~\cite{gdp}.
State change comparison of multiple data items is often necessary for quantitative analysis in various application domains~\cite{izquierdoevaluation, de2012referral, fiori2018economic, bonomi2013weight},
such as sociology, medical science, finance and biology..
% , such as social science, financial engineering, medical science and biological science.
For instance, the NBA league will assess all players’ improvements compared to the previous season and give the NBA Most Improved Player award to the player with the biggest progress~\cite{mahmood2021using}.
During the COVID-19 pandemic, social media often show the changes of daily new case numbers of different countries/regions between today and yesterday to indicate the pandemic trend~\cite{lin2020estimating}.

Despite its wide usage and significant importance, it is a non-trivial task to achieve effective state change comparison via data visualizations.
The challenges mainly originate from two aspects.
First, state change comparison often involves multiple data items (\textit{e.g.}, multiple NBA players and different countries/regions in the above examples).
With an increase of data items, visual clutters can easily appear and affect the effectiveness of visual comparison by human users.
Second, apart from state changes themselves, both the initial and final states of different data items are necessary to provide a context for such state change comparisons in real applications.
However, it remains unclear on enabling an easy comparison of multiple state changes while preserving the context.

%%% Part 2: Existing techniques for state change comparison and their major issues

% Add the references of existing papers on visualization comparison tasks.
% However,
% despite its wide usage and significant importance,

\begin{figure}[t]
\centering
\includegraphics[width=0.9\columnwidth]{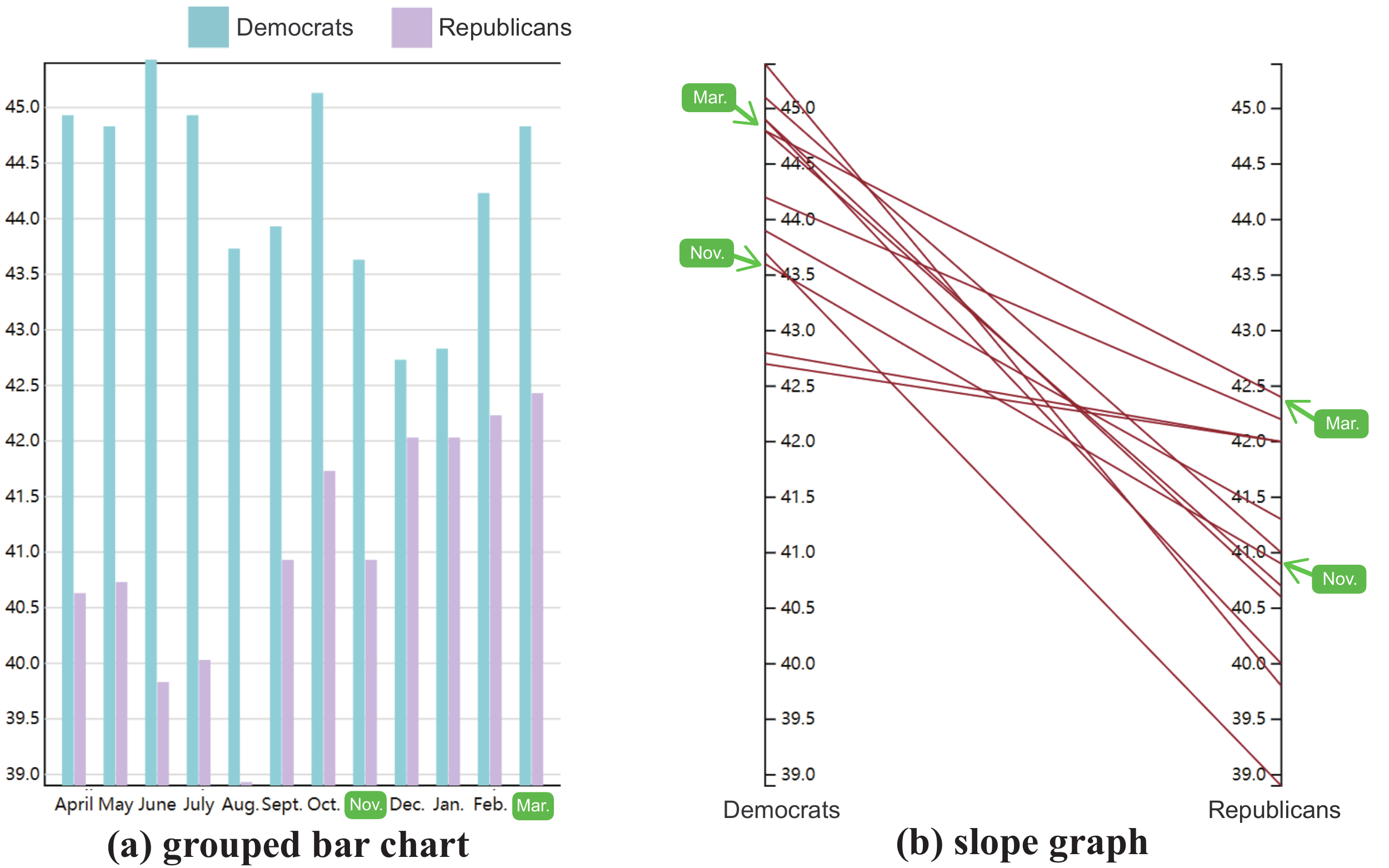}
\caption{
Two visualizations for the comparison of the approval rate of democrats and republicans across the last 12 months. (Access date: 2022-6-19). Grouped bar charts (a) and slope graphs (b) cannot support a quick and accurate comparison of the approval rate difference in target months highlighted by green annotations.
% encode two data series by juxtaposed bars and the difference is encoded by the bar height differences. (b) Slope graphs visualize data series with the positions of line end points, encoding the difference by the line slopes.
}
\label{fig:1}
\end{figure}

Comparison is an important visualization task~\cite{brehmer2013multi, schulz2013design, wehrend1990problem, srinivasan2018s} and various visualization techniques have been developed for visual comparison of different types of data, such as trees~\cite{munzner2003treejuxtaposer,holten2008visual,procter2010visualization} and  flow fields~\cite{pagendarm1995comparative, verma2004comparative, urness2006strategies}.
% \yong{Add some other types of papers here?}
However, little research has been done on effectively visualizing and comparing multiple state changes. According to our prior survey~\cite{ruan2021intercept} and observations from existing studies~\cite{srinivasan2018s}, grouped bar charts and slope graphs are often used for visual comparison of state changes due to their simple visual design and easy implementation.
% where both state changes and the initial and final states of data items can be intuitively visualized.
Grouped bar charts (Fig. \ref{fig:1}(a)) often use two bars within the same group to indicate the initial and final states of a data item and leverage their height difference to implicitly indicate the state change.
Instead, slope graphs (Fig. \ref{fig:1}(b)) visualize state changes as the \textit{slopes} of line graphs, where the initial and final states of data items are shown on two vertical axes.
Despite the simplicity and prevalence of group bar charts and slope graphs,
they suffer from two major issues that affect their effectiveness for visual comparison of state changes.
The first major issue stems from their visual encoding choices.
Group bar charts rely on \textit{height differences of bars} to visualize state changes, but existing perception research on group bar charts ~\cite{burns2009modeling} has demonstrated that people perform badly in comparing height differences of grouped bars.
Slope graphs employ \textit{slope} to indicate the magnitude of state changes.
However, prior studies have shown that slope is a less accurate visual encoding than other visual encodings (\textit{e.g.}, \textit{length})~\cite{cleveland1987graphical,cleveland1986experiment,cleveland1985graphical}.
Their second major issue is visual scalability. With the increase of data items, the bars of grouped bar charts will become thin and difficult to recognize~\cite{eick2002visual} (Fig. \ref{fig:1}(a)) due to the limited screen space, and visual comparison of state changes is also distracted by various short and tall bars~\cite{talbot2014four}. 
For slope graphs, serious crossings and visual clutters will appear (Fig. \ref{fig:1}(b)), making it hard (if not impossible) to compare state changes.
When the number of data items exceeds a certain threshold that slope graphs and grouped bar charts can handle (as shown in Figs. \ref{fig:case2}(b) and (d)), users may even be forced to choose data tables alternatively visualize state changes of data items~\cite{willers2017methods}.

%% Part 3: Our approach and how they address the limitations of existing techniques
To fill the research gap, we propose a novel radial visualization, \toolName{}\footnote{This paper is extended from our IEEE VIS 2021 short paper~\cite{ruan2021intercept}.}, to facilitate an effective context-aware comparison of state changes.
As shown in Fig.~\ref{fig:3}c, the proposed visualization consists of two circular axes to indicate the initial and final states of multiple data items.
As will be rigorously proved in Sec.~\ref{sec:proof}, the \textit{lengths of line segments} intercepted by the inner axis explicitly reflect the state changes of different data items.
In addition, it allows users to interactively adjust the radius of the inner axis and filter the data items with larger state changes smoothly, which reduces visual clutters and helps users focus on the relatively more important large state changes. The challenge of comparing relatively-similar state changes can also be mitigated in the process of radius adjustment of the inner axis, as \toolName{} intrinsically enables magnification of differences among similar state changes.
We present two usage scenarios to show how our approach can be applied to analyze specific state change comparison tasks. 
We also conducted quantitative metric evaluations and a well-designed crowd-sourcing user study to demonstrate the effectiveness and usefulness of \toolName\ in comparison with the baseline approach.

%   - Radial design
%     - Visual encoding by length
%     - Smooth interactions or mental-map preserving filtering
%     - Preserve the contexts
    
%   - Open-source the code

%% Part 4: Contributions

The major contributions of this paper can be summarized as follows:

\begin{itemize}
    \item We proposed \toolName, a novel radial visualization, to facilitate an effective context-aware comparison of state changes. It leverages the length of line segments to explicitly encode state changes. Also, it allows users to interactively adjust the inner axis radius, enabling quick and smooth filtering of large state changes and easy identification of the differences between relatively-similar state changes.
    % , which has a more accurate graphical perception than line slopes and bar height differences.

    % \item We proposed two useful features, \textit{\textit{i.e.}}, filtering and comparison, based on the interaction with the inner axis. 
    % The filtering feature can support quick and smooth filtering for large state changes, while the comparison feature enables a promoted comparison for two close differences, making the users identify the relationships of the differences more accurately and confidently.
    
    \item We extensively evaluated \toolName{} through two usage scenarios, quantitative metric evaluations, and well-designed crowd-sourcing user studies. The results demonstrated the usefulness and effectiveness of \toolName{} for state change comparison.
    
    % \item We present usage scenarios to demonstrate the usefulness of \toolName. Meanwhile, to systematically evaluate the effectiveness and usefulness of \toolName, we conduct the metric evaluation and user study in comparison with the baseline approach.
\end{itemize}

We have implemented our approach as a publicly-available JavaScript package called \textit{interceptgraph}~\footnote{\url{https://www.npmjs.com/package/interceptgraph}}, benefiting common users who need to conduct state change comparisons.
Also, the online demo, source code and implementation examples of \toolName\ can be accessed here: \textcolor{blue}{\url{https://interceptgraph.github.io/}}.

\section{Related Work}

Our work is relevant to prior studies on visualization for comparison tasks and radial visualizations, which will be discussed in this section.

% In this section, we explore related work on
% visualizations for comparison
% % graphical perception,
% and radial visualization.

\subsection{Visualization for Comparison Tasks}

For various visualizations for comparison tasks, Gleicher~\cite{gleicher2017considerations} proposed a taxonomy that categorized the visual designs for comparison tasks into three groups, \textit{i.e.}, \textit{Juxtaposition}, \textit{Superposition}, and \textit{Explicit encoding}.
Juxtaposition designs show two data series to be compared separately, visualizing the differences via graphics or views next to each other. 
% According to the appearance of the design, they can be further grouped into dual-views~\cite{namata2007dual} or side-by-side views~\cite{lam2007overview}.
% Aside from the space, juxtaposition will occur in time as well~\cite{gleicher2011visual}.
One of the earliest examples of juxtaposition comparison was displayed by the English Hexapla New Testament in 1841.~\cite{ts1841}.
After that, a set of novel juxtaposition designs are proposed for domain-specific comparison tasks. For example, Albers \textit{et al.}~\cite{albers2011sequence} proposed juxtaposed views called Sequence Surveyor to highlight outliers of genomic sequences. Sherlock~\cite{white2004sentence} and Turnitin~\cite{sutherland2005turnitin} introduced a tool to help teachers find plagiarisms via side-by-side comparison.
Also, the common-used visualization for comparison tasks~\cite{srinivasan2018s}, \textit{i.e.}, grouped bar chart, reflects state changes by juxtaposed bars for general users.
Superposition designs highlight the difference between multiple states in the same space. 
% The most straightforward approach for superposition design is making one object semi-transparent to one another~\cite{shaw1999interactive}.
Other techniques were also proposed for superposition designs, such as the union graph approach. For example, 
% Gevol~\cite{collberg2003system} highlights the evolution of software using large graphs, 
% \yong{Pls check my Chinese comments. Also, pls exhaustively check all the related work and see if the same issues appear in other parts.}
Jianu \textit{et al.}~\cite{jianu2009visual} utilized superposition visualizations for the protein-protein interaction network analysis.
Designs of explicit encoding directly visualize the differences (or relations) between objects. 
% The most straightforward examples are single bar charts and difference charts. Besides, 
% Mizbee~\cite{meyer2009mizbee} explicitly displays the relationships between two genomes.
Darling \textit{et al.}~\cite{darling2004mauve} introduced a visual comparison approach for highlighting genomic DNA in the presence of rearrangements and horizontal transfer. One approach to increase the scalability of widely-used grouped bar charts is to depict the difference between objects directly using explicit encoding, resulting in the loss of the context data values~\cite{srinivasan2018s}.

Compared with the above visualizations, our design \toolName\ is a novel visualization that can be applied for a wider range of usage scenarios for domain-agnostic users.
According to the taxonomy by Gleicher~\cite{gleicher2017considerations}, our visualization \toolName\ can be categorized into \textit{juxtaposition} designs, which show two-series data values separately. However, we address the key challenge in juxtaposition design (\textit{i.e.}, it is difficult to highlight the relationships between separate objects~\cite{gleicher2011visual}) by a line segment encoding.

\subsection{Radial Visualizations}

Hoffman \textit{et al.}~\cite{hoffman1997dna} first introduced the term \textit{radial visualization} in 1997. 
Draper \textit{et al.}~\cite{draper2009survey} categorized the radial visualizations into three groups, \textit{i.e.}, \textit{polar plot}, \textit{space filling}, and \textit{ring}.
Polar plot visualizations refer to those radial graphics whose center is the focus of the whole chart. 
% Based on the hierarchy and branches around the center, the polar plot could be further divided into tree patterns and star patterns. 
A typical polar visualization is tree visualizations~\cite{teoh2002rings, reggiani1988proposed, carpano1980automatic}, which can be used to view hierarchical data structure, while other variants of polar visualization, \textit{i.e.}, the star visualizations\cite{spence2001information,hetzler1998multi,havre2001interactive}, are for the ranking of the search results.
Space filling radial visualizations can be categorized into three types, \textit{i.e.}, concentric, spiral, and Euler~\cite{yang2002interring}. The concentric~\cite{cugini1996interactive} and Euler~\cite{hong2003zoomology,van2003bubbleworld} types can be used to browse hierarchical data or relationships among disparate entities. The spiral types~\cite{weber2001visualizing,dragicevic2002spiraclock,carlis1998interactive}, however, are used for viewing serial periodic data, such as time-based data, due to their characteristics of spiral-shaped glyphs emanating from their origin.
Ring-based radial visual designs can be divided into connected and disconnected ring types. Specifically, connected ring types~\cite{livnat2005visualization} contain the nodes positioned around the circumference of the ring which is connected by a set of line segments, while disconnected ring types~\cite{suntinger2008event,draper2008votes} are with additional nodes optionally appearing in the ring’s interior.
More recently, radial visualization is an active topic in the visualization community. For example, Shi \textit{et al.}~\cite{shi2018novel} introduced a tool to identify the potential attack in intrusion detection systems. Long \textit{et al.}\cite{van2015optimal} studied an algorithm for maximizing the quality of radial visualization for classifier data.
% Additionally, prior studies further discussed the strengths and weaknesses of radial visualization through various methodologies~\cite{diehl2010uncovering,goldberg2011eye}.

According to the taxonomy presented by Draper \textit{et al.}~\cite{draper2009survey}, \toolName\ belongs to
% the subtype Connected Ring Pattern under Ring Based,
ring-based radial visualizations.
% \yong{1. What is the point of mentioning the sub group? 2. Without any explanation, nobody knows what the subtype mean.}
% which uses line segments to connect nodes denoting dimension members of states. Accordingly, 
\toolName\ preserves the advantages of radial visualization and further extends static radial methods via flexible interactions, making it able to compare items more accurately and smoothly.

\section{Intercept Graph}

In this section, we introduce the visual design of \toolName, and the adjustment of the inner axis.
% The examples of \toolName\ can be accessed via the URL: \textcolor{blue}{\url{https://interceptgraph.github.io/}}.

\subsection{Visual Design}
\label{subsec:visual_design}

\toolName\ consists of inner and outer circular axes and each data item is represented by a line segment connecting the inner axis and outer axis.
Also, the line segments intercepted by the inner axis allow a quick filtering and accurate comparison of state changes.
% components, \textit{i.e.}, semi-circular axes and a set of line segments connected from the inner axis and the outer axis.
% We also considered two design alternatives for \toolName.

\begin{figure}[t]
\centering
\includegraphics[width=\columnwidth]{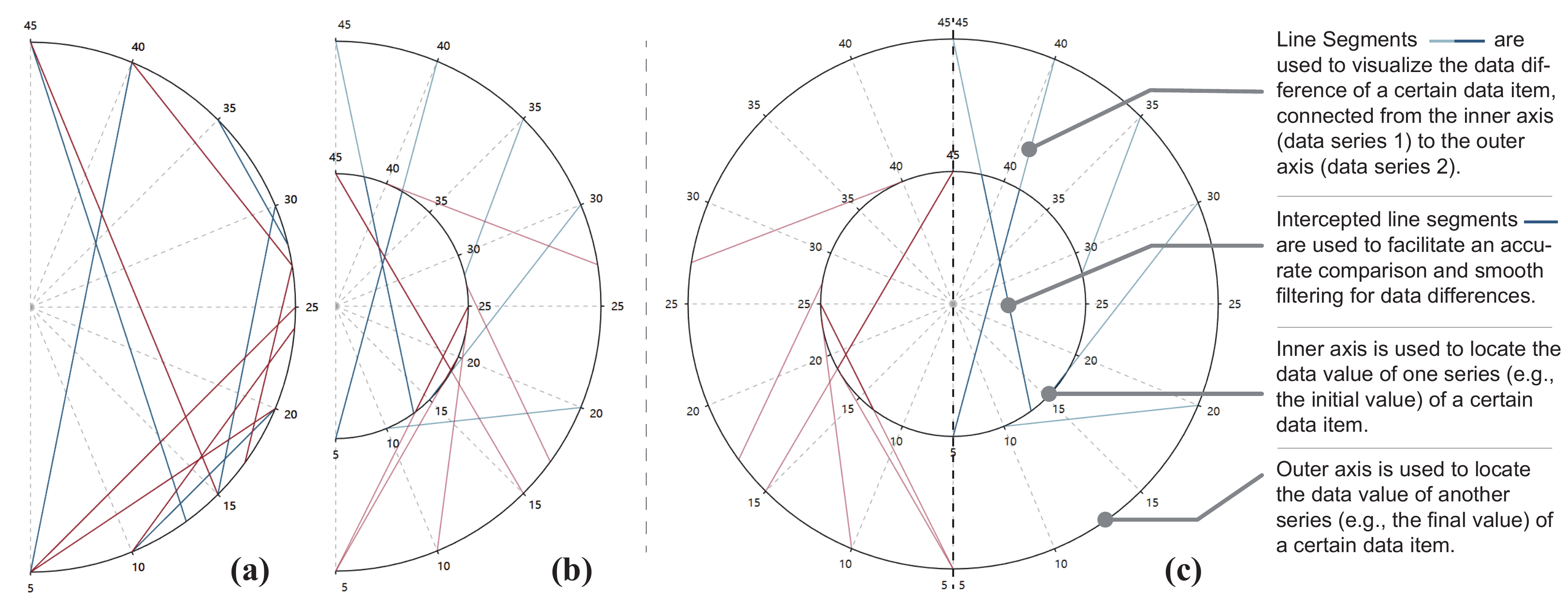}
\caption{Visual design of \toolName. (a) (b) Design alternatives of \toolName. (c) The final design of \toolName, which supports the smooth filtering and accurate comparison while mitigating the visual clutter of (a) and (b).}
\label{fig:3}
\end{figure}

\textbf{Semi-circular axes.} 
As shown in Fig. \ref{fig:3}(c), the outer axis indicates the outer semi-circular axis used to locate one series of data items (\textit{e.g.}, initial state), while the inner axis indicates the inner semi-circular axis used to locate another series of data values (\textit{e.g.}, final state).
% the outer axis and the inner axis are formulated by semi-circles, which are used to locate the two series of data values.
The outer axis has a fixed radius, while the radius of the inner axis can be interactively adjusted by users and is always no greater than that of the outer axis.
The inner and outer axis are both linear scales.
% To make the most of space within the axes for comparison,
To make full use of the limited space and facilitate easy comparison of state changes,
% the ranges of the inner and outer axis are both the extent of the union of two-series data values instead of starting from zero.
the ranges of the inner and outer axis are kept the same and set as the minimum and maximum values of all the data items.
The whole design is divided into two parts, \textit{i.e.}, the left semi-circular axis and the right semi-circular axis. Specifically, the right semi-circular axis is for data items with non-negative state changes, while the left semi-circular axis is for those with negative state changes.
%
%
%
% The directions of the increasing values of the axes on the right half are both clockwise, while the axes on the left half are counterclockwise. \yong{What are you talking about in the commented sentence here?}
%%
%
%
% \yong{Some core information of the visual design is missing: 1) what does the left and right semi-circle represent? 2) Do we use different colors for decreasing and increasing data items? 3) We may still follow the style of the short paper and add a dividing line in Fig. 2c to highlight that the left and right semi-circles have different meanings.}

\textbf{Line segments.}
We leverage line segment to represent the data item. 
The length of line segments can reflect the state changes of different data items and enable an effective comparison between different state changes. The rigorous mathematical proofs will be presented in Sec. \ref{sec:proof}.
% We present the mathematical proofs about this encoding in Sec. \ref{sec:proof}. 
% \yong{The same statement is mentioned twice in the same paragraph!}
%
%
For example, assume that there is a student with mid-term grades of 60 and the final grades of 85, then we can draw a line segment from 60.0 on the inner axis to 85.0 on the outer axis to represent the grade change of 25.0.
% For example, assume we have a controlled trial where one result without an independent variable is 12.0, while the result with the variable is 20.0,
% As mentioned above, to mitigate the visual clutter, we separate the data items based on the signs of the differences. Specifically, data values with a non-negative difference will be placed on the right component, while those with a negative difference will be on the left component.
% To ensure the same luminance for positive and negative state changes,
To differentiate positive and negative state changes, 
the line segments in the right and left parts of the visualization are colored in blue and red, respectively.
% We use blue instead of green to depict the ``rising'' trend with a higher visual comparison efficiency~\cite{javed2010graphical}.
% The difference quantity is mapped to the central angle subtended by the chord linearly, and 
Specifically, a whole line segment has two parts: the intercepted line segment within the inner axis and the line segment between the inner and outer axis. We leverage the intercepted line segments to enable filtering and comparison of state changes. 
We introduce the detailed usage of filtering and comparison features in Section \ref{subsec:adjust_axis}.
To highlight the intercepted parts of line segments, the opacity of the filtered line segments (within the inner axis) is set as 100\%, while the opacity of the line segments filtered out (the area between the inner and outer axis) is set as 30\%.
%
%

% \yong{Pls check my Chinese comments.}

\textbf{Design alternatives.}
Before finalizing the current visual design,
we also considered other two design alternatives for the visualization for comparison tasks.
Fig. \ref{fig:3}(a) shows the initial design of \toolName. We use a single semi-circular axis to locate two data series simultaneously. However, this kind of straightforward design cannot support any filtering or comparison of the changes of two data series. Fig. \ref{fig:3}(b) addresses the limitation with an inner axis component. Through the adjustment of the inner axis, it enables users to filter and compare individual state changes interactively. Through our iterative discussion, we found that it is difficult to observe the visualization with positive and negative state changes simultaneously due to severe visual clutter. Thus, we came up with the final visual design as shown in Fig. \ref{fig:3}(c), which mitigates the clutter by separating data items with rising and dropping trends into right and left components respectively.

\subsection{Interactive Adjustment of Inner Axis}
\label{subsec:adjust_axis}
% To make \toolName\ intrinsically support comparison-related features, we proposed an inner axis design embedded within the outer axis.
%
% Through the adjustment of the inner axis, \toolName\ can facilitate the \textit{filtering} and \textit{comparison} of state changes of a set of data items.
\toolName\ allows users to interactively adjust the inner axis and facilitates smooth \textit{filtering} and accurate \textit{comparison} of state changes of multiple data items.

\begin{figure*}[t]
\centering
\includegraphics[width=\linewidth]{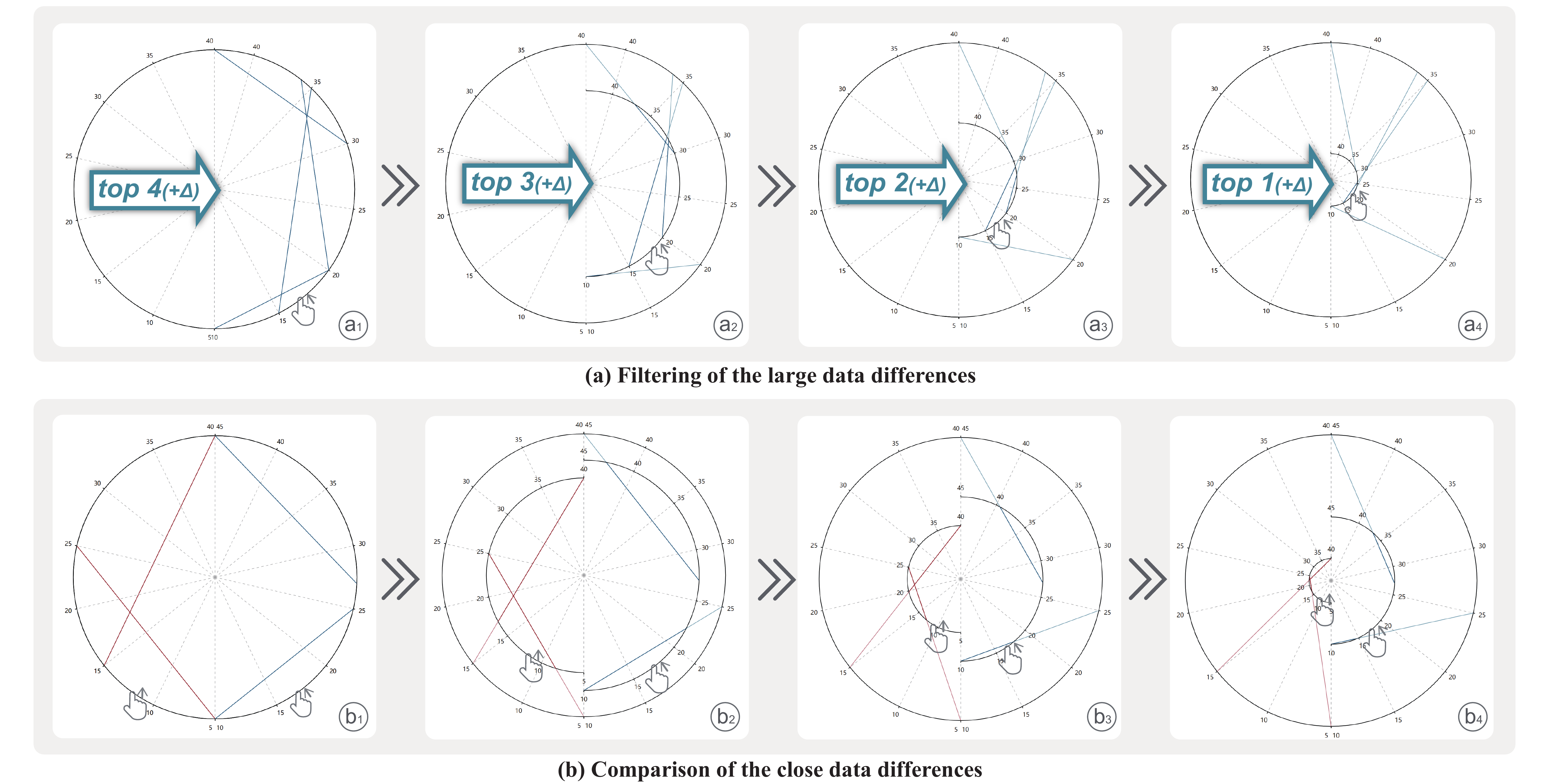}
\caption{The filtering and comparison features of \toolName. (a) illustrated the filtering process of \toolName. The number $n$ in the left half represents the number of the filtered items for each figure. ``$+\Delta$'' indicates the state changes to be filtered are generally positive.
(b) shows the comparison feature of \toolName. The figure on the left indicates \toolName\ that the comparison of two pairs of data items' state changes is difficult to identify. As the inner axis shrinks inward, the relationships of each pair of data items are getting more and more apparent to identify.}
\label{fig:2}
\end{figure*}

\textbf{Filtering.}
Building upon the geometry of \toolName\ that will be illustrated in Sec. \ref{sec:proof}, our approach can support a filtering feature based on the adjustment of the inner axis. Specifically, shrinking the inner axis inward will filter those data items with relatively larger state changes. In the beginning, if the inner and outer axis coincide (\textit{i.e.}, the maximum radius of the inner axis), no data items will be filtered out, and the entire set of data items will be enclosed within the inner axis (Fig. \ref{fig:2}(a\textsubscript{1})); as the inner axis shrinks inward, those data items with larger state changes are still kept within the inner axis and those with the smaller state changes will be filtered out and left between the inner and outer axis. The more the inner axis shrinks, the larger state changes will be filtered within the inner axis (Figs. \ref{fig:2}(a\textsubscript{2}), \ref{fig:2}(a\textsubscript{3}), and \ref{fig:2}(a\textsubscript{4})). Thus, the filtering can be supported by the interaction with the inner axis, enabling users to inspect the filtering process more smoothly.

\textbf{Comparison.}
\toolName\ can intrinsically augment the comparison of a pair of state changes, especially for those data items with similar state changes. As shown in Fig. \ref{fig:2}(b), the difference of the pair of line segments' length can be magnified by shrinking the inner axis inward. More specifically, if the inner axis coincides with the outer axis, there will be no magnification for the length difference (Fig. \ref{fig:2}(b\textsubscript{1})). By shrinking the inner axis inward, the length difference will be magnified gradually until it exceeds the Just Noticeable Difference (JND)~\cite{hecht1924visual} which enables humans to make confident identification (Fig. \ref{fig:2}(b\textsubscript{4})). The more the inner axis shrinks, the larger the difference of a pair of line segments will be magnified. We also provide a rigorous mathematical proof of the comparison feature in Sec. \ref{sec:proof}.

\toolName\ is so named because major features, \textit{i.e.}, filtering and comparison for state changes, are supported by the interaction with the intercepted line segments.
% In addition to the basic visual component introduced above, \toolName\ offers flexible interactions to facilitate better inspections of the state change in the data items.
To enable the users to perform the above \textit{filtering} and \textit{comparison} smoothly, the radius of the inner axis can be adjusted via \textit{dragging}. 
% By dragging the inner axis, the user can perform the filtering and comparison directly on the visual components instead of embedding other components (\textit{e.g.}, slider bar).
Note that \toolName\ will initially be created with a default radius. We define the default radius as $1/2$ of the outer axis radius.
% However, the users can interact with the inner axis at any time if further filtering or comparison is needed. 
The display of the line segments' name labels can be triggered via the mouse hover.
\section{Mathematical Proofs}
\label{sec:proof}

In this section, we prove the visual designs and the major features of \toolName\ (\textit{i.e.}, filtering and comparison) via mathematical proofs. 
% In particular, first, 
% we prove that the length of the line segment is positively correlated with the state change.
% Second, we prove that the length of the intercepted line segment is positively correlated to the state change.
% Third, we prove that the ratio of the two state changes' length can be magnified as the inner axis shrinks inward.

\textbf{Proposition 1:} 
% The line segment connected from the inner axis to the outer axis can reflect the state change between the two values of a certain data item, \textit{i.e.}, \textit{
The length of the line segment is positively correlated with the difference magnitude.

\underline{Proof:}
Given the constant outer axis radius $R$ and inner axis radius $r \in [0,R]$. Assume a certain data item with two values $x$ and $y$, and the difference $d=|x-y|$. Since the inner and outer axis is on a common linear scale and two values $x$ and $y$ are located based on the angles and the radius of the axis, the center angle subtended by the line segment of data item $\alpha \in (0, \pi]$ is proportional to the difference $d$. Hence

$$
    d \propto \alpha.
$$

Applying the Law of Cosines to the length $L$ of the line segment connecting two points, yields

$$
    L = \sqrt{R^2 + r^2 - 2\cdot R \cdot r \cdot \cos \alpha},
$$

for a given $r$, in view of $d \propto \alpha$, this implies the length $L$ is positively correlated with the difference $d$ as desired.

\textbf{Proposition 2:} 
% The difference can be further visualized by the line segments intercepted by the inner axis, \textit{i.e.}, \textit{
The intercepted line segments are positively correlated with the state changes.

\underline{Proof:}
Given the triangle formed by the intercepted line segment and two radii is an isosceles triangle. Thus, the length $L$ of intercepted line segments satisfies
\begin{equation}
\label{equa:1}
        L(x) = 2 \cdot r \cdot \sin x,
\end{equation}

where the angle $x \in [0, \frac{\pi}{2}]$ is the angle between the radius of the inner circle and the vertical line from the center of the circle to the intercepted line segment.

Based on the geometric relationships between the two triangles containing the above vertical line, the angle $x$ can be represented by the angle $\alpha$ and the inner radius $r$, so that

\begin{equation}
\label{equa:2}
        x = \alpha - \arctan{\frac{\frac{R}{r} - \cos{\alpha}}{\sin{\alpha}}}.
\end{equation}

By applying Equation \ref{equa:2} to Equation \ref{equa:1}, we can calculate $L(\alpha)$ as follows:

$$
L(\alpha) = 2 \cdot r \cdot \sin({\alpha - \arctan{\frac{\frac{R}{r} - \cos{\alpha}}{\sin{\alpha}}}}).
$$

% For a given $r \in [0, R]$, the derivative of $L(\alpha)$ can be represented as

% \begin{equation}
% \label{equa:4}
%         L'(\alpha) = 2 \cdot \frac{R^2}{r} \cdot \frac{\frac{R}{r} - \cos{\alpha}}{[\sin^2{\alpha}+ (\frac{R}{r}-\cos{\alpha})^2]\cdot \sqrt{1+(\frac{\frac{R}{r}-\cos{\alpha}}{\sin{\alpha}})^2}}
% \end{equation}

For a given $r \in [0, R]$, the gradient $L'(\alpha)$ is positive given the $\alpha \in (0, \pi]$. Thus, the length $L(\alpha)$ is positively correlated with $\alpha$. Since $d \propto \alpha$, we conclude that the length of the intercepted line segment $L(\alpha)$ is positively correlated with the difference $d$.

Based on the positive correlations, users can identify the state changes based on the length of the intercepted line segments,  
which means that the filtering of the state changes based on the length of the intercepted line segments is also proved.

\textbf{Proposition 3:} 
% When the inner axis shrinks inward, the difference relationships can be identified more easily. In other words, 
The length difference between two intercepted line segments is negatively correlated with the radius of the inner axis.

\underline{Proof:}
Given two line segments with their corresponding center angle $\theta_1$, $\theta_2$ ($\frac{\pi}{2} > \theta_1 > \theta_2$) and the inner axis radius $r \in (R\cos{\theta_2},R]$ ($R\cos{\theta_2}$ is the radius when the line segment with a smaller difference is filtered out), the length of the line segment satisfies

$$
        L_{i}(r) = 2 \cdot r \cdot \sin({\theta_i - \arctan{\frac{\frac{R}{r} - \cos{\theta_i}}{\sin{\theta_i}}}}),
$$

where $i \in \{1,2\}$. Thus, the differences of two given line segment lengths can be represented as

$$
        \Delta L(r) = L_1(r) - L_2(r).
$$

For the inner axis $r \in (R\cos{\theta_2},R]$, the derivative of $\Delta L'(r)$ is identically negative. Thus, the difference in lengths of two intercepted line segments becomes larger as the inner axis radius decreases.

\section{Usage Scenario}
\label{sec-usage-scenario}

In this section, we describe two usage scenarios to demonstrate the usefulness of \toolName\ compared with the existing visualization approaches, \textit{i.e.}, slope graphs and grouped bar charts.

\subsection{Scenario \uppercase \expandafter{\romannumeral 2} - Basketball Players' PPG Ranking Changes}

\begin{figure}[t]
\centering
\includegraphics[width=\columnwidth]{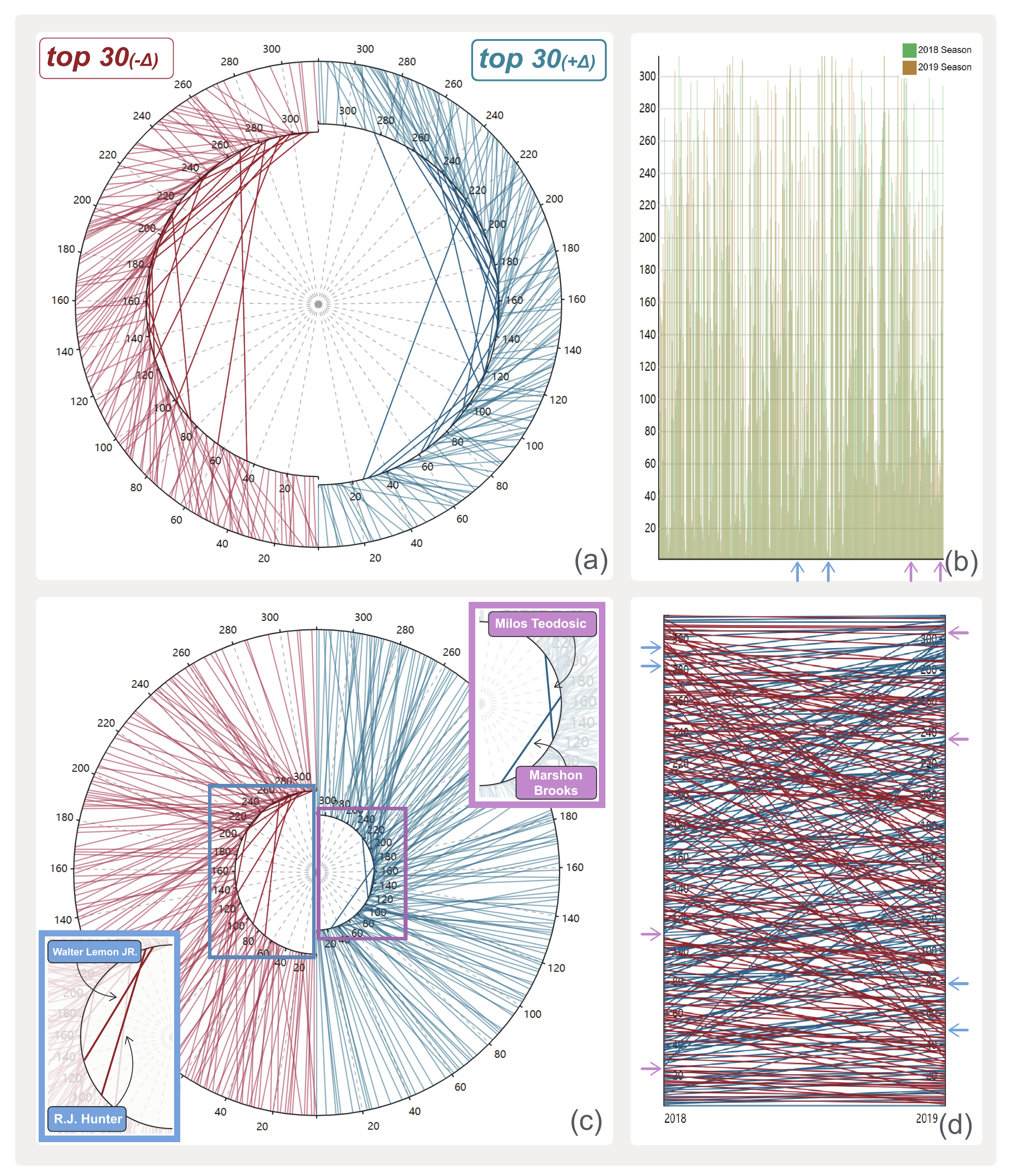}
\caption{Comparative visualizations using a basketball dataset. (a) and (c) is \toolName, where (a) shows the filtering of players with top 30 rising and dropping rankings, and (c) illustrates the accurate comparison of two pairs of players with close PPG ranking differences. (b) and (d) is the slope graph and grouped bar chart respectively, both driven by the basketball dataset which includes over 300 players.}
\label{fig:case2}
\end{figure}

We demonstrate the usefulness of \toolName\ using another National Basketball Association (NBA) dataset, which contains 321 NBA players' Points per Game (PPG) rankings in Season 2018 and 2019~\cite{nbadataset}. The players are those active players with valid PPG records in Seasons 2018 and 2019. We adopt this application for players' statistic comparison because the assessment is of great importance in the NBA, which is the motivation of the foundation of the annual award Most Improved Player (MIP)~\cite{fearnhead2011estimating}.
For the basketball dataset, the data item is the individual player, and the two states are the rankings of each player's PPG in Seasons 2018 and 2019 respectively. As shown in Fig. \ref{fig:case2}, we plot the data using three visualizations (\textit{i.e.}, \toolName\ (Fig. \ref{fig:case2}(a)), grouped bar chart (Fig. \ref{fig:case2}(b), and slope graph (Fig. \ref{fig:case2}(d))).
% For the slope graph and grouped bar chart, it is apparent that the two conventional visualization approaches both suffer from a severe visual clutter with 321 data items. Specifically, the slope graph faces a significant line crossing and occlusion even though we differentiated the signs of state changes with different colors.
% It is so difficult to identify a line segment and perform a further comparison.
% Similarly, for the grouped bar chart, 

% \textbf{Coarse-grained filtering for MIP candidates.}
\textbf{Initial filtering for MIP candidates.}
For the grouped bar chart (Fig. \ref{fig:case2}(b)), we can not even identify the height of each bar due to the severe visual clutter. The number of data items (\textit{i.e.}, over 300) has already exceeded its visual scalability (\textit{i.e.}, a few dozens)~\cite{eick2002visual}, making each bar too thin to be inspected and compared. Further identification of the bar height difference seems quite burdensome for human eyes.
% For slope graph (Fig. \ref{fig:case2}(d)), the visual clutter has slightly improved because we can identify the individual line slopes although with great effort. However, we still feel difficult to compare the line slopes due to severe line crossings and occlusion.
Compared with the grouped bar chart, we can identify individual line slopes with efforts from the slope graph (Fig. \ref{fig:case2}(d)), but it is still difficult to compare line slopes due to severe line crossings and occlusions. 
% We attempt to leverage \toolName\ to address the issues encountered in the slope graph and grouped bar chart. 
Instead,
\toolName\ mitigates the visual clutter significantly, as shown in Fig. \ref{fig:case2}(a).
To filter the 30 MIP candidates out of 321 players, we can drag the inner axis on the left inward to exclude those players with relatively small PPG changes. We stop dragging when there are about 30 intercepted line segments left within the inner axis (Fig. \ref{fig:case2}(a)). 
We find that many excellent players have a ranking increase of over 100 compared with Season 2018. 
After hovering on the line segments to observe the player's name, JaKarr Sampson and R.J. Hunter even made a rank move-up of over 210, as indicated by the longest intercepted line segment near the circle center.
% \yong{How can we know that from the figure?! The figure does not show any names!}
After the selection of players with much progress, we are also curious about who got worse in the league compared to Season 2018. So we further drag the inner axis on the right until there are about 30 line segments. Through the smooth interaction and mouse hovering, we can quickly find the 30 players who got worse most in the league (\textit{e.g.}, Marshon Brooks, Milos Teodosic, etc.).
% \yong{Again, how can we know that from the visualization?! }
This finding is confirmed by the sports news that most of these players suffered serious injuries in Season 2019, leading to worse performance on the court~\cite{injury}.

\textbf{Detailed comparison of MIP candidates.}
After filtering the MIP candidates, we conduct further comparisons among the filtered candidates.
First, we try to compare the target line segments in detail using the slope graph, as highlighted by the purple and blue arrows in Fig. \ref{fig:case2}(d). However, we cannot even identify the two line segments we tend to compare due to the significant visual clutter, making it much tougher for an accurate comparison of the slopes of target line segments.
For the grouped bar chart (Fig. \ref{fig:case2}(b)), the inspection seems to be even more difficult. Even if we can figure out the target objects to be compared (as indicated by the arrows in Fig. \ref{fig:case2}(b)), it seems impossible for us to compare the bar height differences due to the bars' thin width and a huge amount of distracting bar groups between the two target bar groups.
We further leverage \toolName\ to address the above issues.
We first drag the inner axis inward, and those line segments with relatively small PPG ranking changes can be easily filtered out and gathered in the area between the inner and outer axis.
% We first drag the left inner axis inward, and those line segments for small differences gathered around the circumference of the outer axis can be easily filtered out.
% % \wys{However, the relationships of all candidates are still not clear enough for determination.}
% % \yong{What are ``the relationships of all candidates''? Make it concrete.}
% However, it is still not clear enough to compare the rank increase of all the candidates.
We then further shrink the inner axis until the relationships of players with the top 3 PPG increase are apparent, as indicated in Fig. \ref{fig:case2}(c). After we hovered on the line segments, it is clear that the player with the largest PPG increase is JaKarr Sampson, as indicated by the longest intercepted line segment.
% \yong{Still not clear. Why not specifying it in the figure?}
Also, as highlighted by the blue rectangle in Fig. \ref{fig:case2}(c), it can be identified confidently that the longer intercepted line segment is R.J. Hunter, while the shorter one is Walter Lemon JR. Similarly, we then compare the two players with the top 2 PPG decreases on the right half. It is easy for us to identify the relationships between two line segment lengths via adjusting the inner axis as highlighted by the purple rectangle in Fig. \ref{fig:case2}(c). It is clear that Marshon Brooks has a larger PPG decrease than Milos Teodosic as indicated by the below longer line segments.

% \textbf{Summary.}
% Given the dataset with over 300 data items, the performances of the three visualizations are different.
% For the filtering of players with large PPG changes, \toolName\ supports smooth filtering for the top 30 MIP candidates, while the slope graph and grouped bar chart cannot support smooth filtering for such a dataset.
% For the comparison of the close difference, \toolName\ magnifies the ratio of the line segment length, making it easy to compare over 300 plus data items. However, the two conventional visualizations face a significant limitation due to the severe visual clutter.
% \yong{Have you checked Prof. Guan's comments?}
\section{metric evaluation}
\label{sec:metric_evaluation}

In this section, we conducted the metric evaluations to demonstrate the effectiveness of \toolName.
According to the usage scenarios illustrated in Section~\ref{sec-usage-scenario}, the performance of state change comparison and visual scalability of the grouped bar chart are generally worse than those of slope graph and \toolName.
% We set the baseline as the slope graph only because the comparison performance and visual scalability of the grouped bar chart are generally worse than that of the slope graph and \toolName\ based on the experiments in usage scenarios.
Thus, we focus on comparing the effectiveness of \toolName\ with slope graphs in terms of their line segment based visual encodings via metric evaluations.
We first introduce the metrics used in our evaluation and then run a quantitative study on the generated dataset.
% using these metrics.
All the experiments were conducted using the default radius of the inner axis without any dragging interactions on the inner axis.

\subsection{Metrics}

We evaluated the visualizations for comparison tasks from two perspectives: \textbf{line crossing} and \textbf{intensity ratio}. 
More specifically, line crossing is a quantitative metric to evaluate the visual complexity of the visualizations consisting of line segments. Previous studies have reported that line crossings significantly impair the readability of the graphs~\cite{sun2018effect}. Alemasoom \textit{et al.}~\cite{alemasoom2016energyviz} agreed that the crossing is a crucial metric for the perception of graphs. 
Inspired by the prior work~\cite{heer2010crowdsourcing,lu2021modeling}, we leveraged the metric of intensity ratio to measure the difference between two stimuli's intensity (\textit{i.e.}, intercepted length of line segments for \toolName\ and line slopes for slope graph).

\textbf{Line Crossing:}
We used the metric line crossing introduced by Purchase~\cite{purchase2002metrics}, which reflects the visual complexity of the edges in a chart.
We applied the line crossing for evaluation because \toolName\ and slope graph both encode data items with line segments, introducing a potential visual clutter by line crossings. The calculation of line crossings of a chart is as follows:

\begin{equation*}
\label{equa:9}
lineCrossing(E) = \sum\nolimits_{p,q \in E,p \neq q} crossing(p,q),
\end{equation*}

where $E$ represents the set of all line segments in the chart. $p,q \in P$ are two line segments in the set $E$. The function $crossing$ will return 1 if two line segments $p,q$ have intersection and 0 otherwise.

\begin{figure*}[t]
\centering
\includegraphics[width=\linewidth]{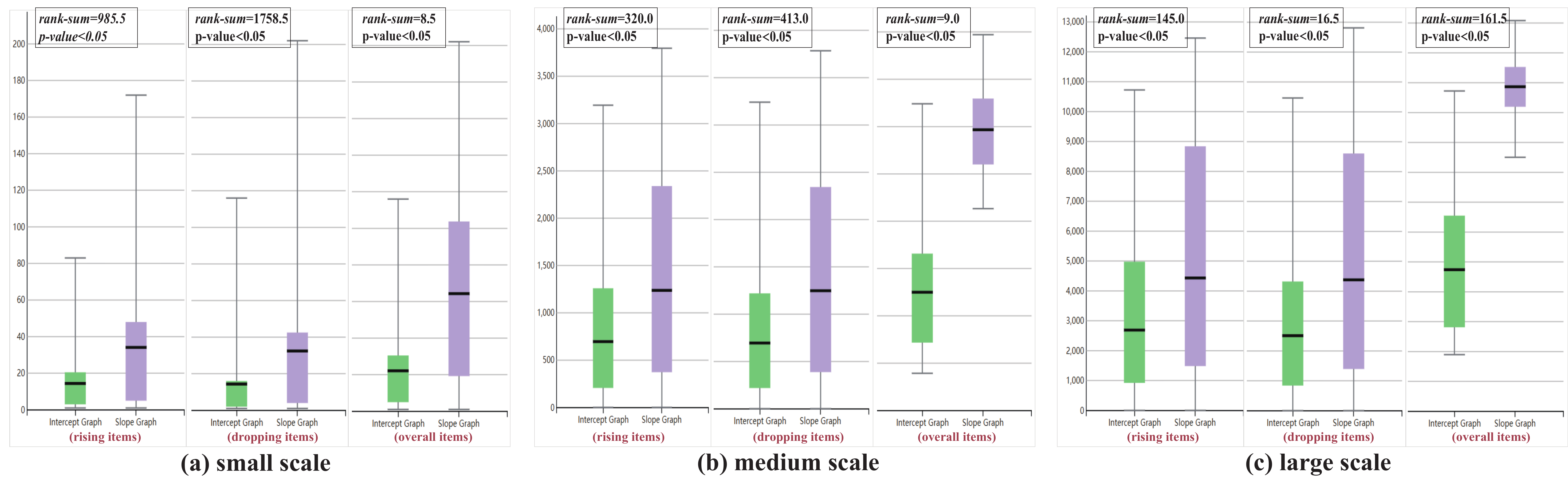}
\caption{Metric evaluation of line crossing on randomly-generated datasets to compare the performance of proposed \toolName\ and the baseline approach slope graphs. The error bars are  95\% confidence intervals. Wilcoxon test statistics are reported at the top left of each figure. The p-values of each experiment are much less than 0.05, indicating a significant improvement in line crossing over the slope graphs.}
\label{fig:4}
\end{figure*}

\begin{figure}[t]
\centering
\includegraphics[width=0.95\columnwidth]{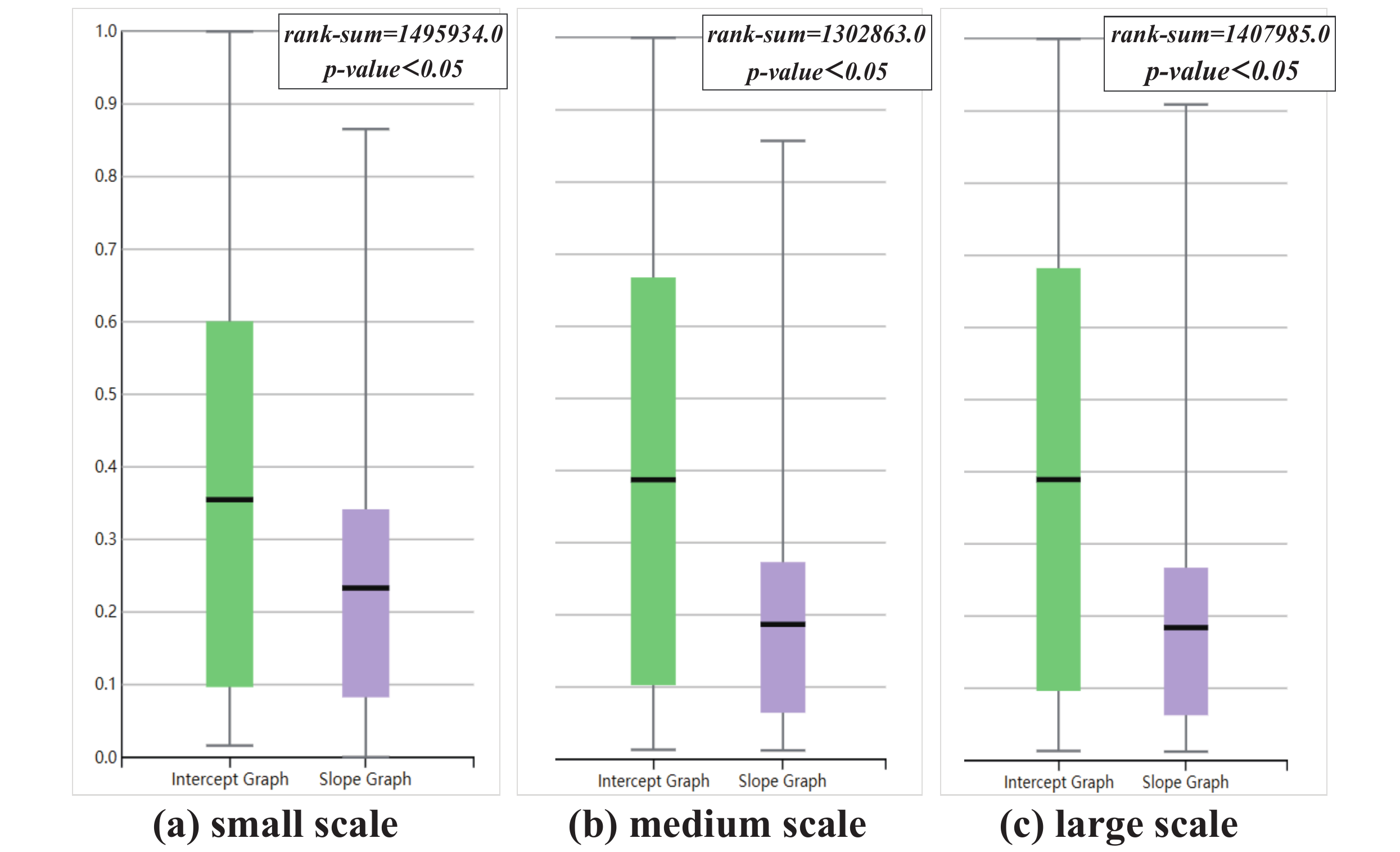}
\caption{Metric evaluation of intensity ratio on randomly-generated datasets to compare the performance of proposed \toolName\ and the baseline approach slope graphs. The error bars are  95\% confidence intervals. Wilcoxon test statistics are reported at the top right of each figure. The p-values of each experiment are much less than 0.05, indicating a significant improvement of intensity ratio over the slope graphs.}
\label{fig:5}
\end{figure}

\textbf{Intensity Ratio:}
We defined the metric intensity ratio to evaluate the relationship between a pair of stimuli's intensity, reflecting how easy/difficult for human eyes to identify the differences between two data items. For \toolName, the intensity is the length of intercepted line segments within the inner axis which the users will use for the difference comparison of two data items; for slope graphs, the intensity is the slope (\textit{i.e.}, the tangent of inclination angle) of the line segments. Basically, the larger the intensity ratio of two data values, the easier for humans to identify the relationships. 
% To narrow our scope on those data item pairs with a close difference, we only focus on the intercepted line segments, because the comparison of two line segments within the inner axis and between two axes is easy for the users to identify based on the line segments' positions and intercepted lengths simultaneously. 
Assume that $I(k)$ indicates the intensity of stimulus $k$, the calculation of the intensity ratio is as follows:

\begin{equation*}
\label{equa:10}
intensityRatio(K)=\frac{1}{K} \sum\nolimits_{p,q \in K,p \neq q} \frac{|I(p)-I(q)|}{max(I(p),I(q))},
\end{equation*}

where $K$ represents the set of selected data items by the intercepted line segments within the inner axis of \toolName. The parameters $p,q \in P$ are two selected line segments in the set $K$. The $intensity$ is the corresponding lengths of \toolName\ and the line slopes of slope graphs. Note that we use the maximum instead of the minimum to avoid an explosion of the ratio for two lengths where the smaller length is close to zero.

\subsection{Metric Experiments}

Given the two metrics, \textit{i.e.}, line crossing and intensity ratio, we generated datasets to evaluate the effectiveness of \toolName. We quantitatively evaluated \toolName\ in comparison with slope graphs. 
% We do not consider the grouped bar charts as our targets because the perception of comparison of the unaligned height is worse than the slopes~\cite{wu2016graphical, cleveland1986experiment, cleveland1985graphical}, in line with the findings through the case studies.

\textbf{Dataset Generation:}
The experiment datasets were generated in the following way. First, we defined three scale levels (\textit{i.e.}, the small-scale, the medium-scale, and the large-scale) of the number of data items. For each scale level, an increment of 5 data items was set to cover a wider range. The item number of each scale level is shown in Table \ref{table:1}.
Assume that the axis ranges of the two data series were from 0 to 100. For each \textit{randomly-generated} dataset, we tried to generate all possible distributions for the given number of data. Specifically, as shown in Table \ref{table:1}, we applied the Gaussian Distribution to the data points generation and set three types of means and three types of standard deviations to data series (\textit{i.e.}, $d_1$ and $d_2$), respectively.
We, therefore, generated 1215 randomly-generated datasets in total: 3 scale levels with 5 scale numbers per level $\times$ 9 Gaussian Distributions of data series $d_1$ $\times$ 9 Gaussian Distributions of data series $d_2$.

\begin{table}[h]
\caption{Parameters for data generation and number of datasets.} 
\centering
\renewcommand{\arraystretch}{1.5}
\begin{tabular}{l|lc}
\hline
     & \multicolumn{1}{l|}{5} & small scale: $\{5, 10, 15, 20, 25\}$                                                                                                                      \\
     \cline{2-3} 
$\times 15$     & \multicolumn{1}{l|}{5} & medium scale: $\{100, 105, 110, 115, 120\}$                                                                                                               \\
     \cline{2-3} 
  & \multicolumn{1}{l|}{5} & large scale: $\{200, 205, 210, 215, 220\}$\\
\hline
$\times 9$   & \multicolumn{1}{l|}{9} & \begin{tabular}[c]{@{}l@{}}data series $d_1\sim N(\mu_1,\sigma_1)$\\
$\mu_1 \in \{25,50,75\}, \sigma_1 \in \{5,10,20\}$\end{tabular} \\
\hline
$\times 9$   & \multicolumn{1}{l|}{9} & \begin{tabular}[c]{@{}l@{}}data series $d_2\sim N(\mu_2,\sigma_2)$\\
$\mu_2 \in \{25,50,75\}, \sigma_2 \in \{5,10,20\}$\end{tabular} \\
\hline
\textbf{1215} &                        & \textbf{total number of datasets}                                                                                                                                       \\
\hline
\end{tabular}
\vspace{10pt}
\label{table:1}
\end{table}

\textbf{Procedures:}
For line crossing measurements, we calculated the line crossing according to the sign of state change separately. 
Specifically, as \toolName\ intrinsically separates positive and negative state changes on the right and left part, while the slope graph draws them in a common area, we split the quantitative experiments into three groups, \textit{i.e.}, the line crossing for the rising data items only, the line crossing for the dropping data items only, and the overall line crossing for all data items. For the overall line crossings of \toolName\, we simply summarized all numbers of line crossings of rising and dropping data; for the overall line crossing of slope graphs, we calculated the line crossings of all line segments regardless of the sign of the differences. We reported the measurement results based on the combinations of the scale level (\textit{i.e.}, the small scale, the medium scale, and the large scale) and the signs of the difference (\textit{i.e.}, rising values, dropping values, and overall values).

For intensity ratio measurements, we randomly selected the same 10 pairs of data items in each \toolName\ and slope graph to calculate the intensity ratio metrics. Last, we reported the measurement results in regard to each scale level (\textit{i.e.}, the small scale, the medium scale, and the large scale). 

We measured the metric scores on the 1215 randomly-generated datasets. We first ran a Shapiro-Wilk test~\cite{shaphiro1965analysis} on each distribution to check for normality. The results show that all measured metrics are not normally distributed. 
Thus, we ran a non-parametric hypothesis test for two paired group comparisons, \textit{i.e.}, Wilcoxon test~\cite{conover1999practical}, for the post-hoc analysis. All the tests were conducted with a standard significance level of $\alpha = 0.05$.

\textbf{Results:}
We reported the metric evaluation results on the randomly-generated datasets as follows:
1) For the line crossing metrics (Fig. \ref{fig:4}), \toolName\ consistently has good performance.
Specifically, for small-scale datasets, \toolName\ is better than the slope graph on both rising, dropping data items, and the overall line crossing.
% meanwhile, \toolName\ performs significantly better than the slope graph in terms of the overall line crossing. 
For medium and large-scale datasets, the line crossing numbers of \toolName\ are nearly a half of that in the slope graph for both rising and dropping data items. Similarly, \toolName\ is significantly better than the slope graph in terms of the overall line crossings.
2) For the intensity ratio metric (Fig. \ref{fig:5}), \toolName\ performs consistently better than slope graphs, especially for the medium and large scale dataset where the magnification of the difference relationships is significantly larger than that of slope graphs.
For both metrics, the p-value of Wilcoxon tests is consistently much less than the error bar (\textit{i.e.}, 0.05), showing significant improvements of \toolName\ in terms of line crossings and intensity ratio over slope graphs.

\section{User Study}

To further evaluate the effectiveness and usability of \toolName, we conducted a carefully-designed user study for comparing \toolName\ with the baseline visualization (\textit{i.e.}, slope graph). 
We only focus on the slope graph because the comparison performance of the grouped bar chart is generally worse than that of the slope graph and \toolName\ according to our findings in usage scenarios.
Specifically, we first conducted a quantitative user study to evaluate the time cost and accuracy of identifying the relationships of differences of a pair of data items. We then conducted a post-study questionnaire to collect the participants' qualitative feedback compared with the baseline approach. 
The details of the online study system are provided in Appendix A.

\subsection{Participants and Apparatus}

We conducted the user study via the crowdsourcing platform, \textit{i.e.}, Prolific\footnote{\url{https://www.prolific.co/}}. We recruited 50 participants (21 female, $age_{mean}=30.32$, $age_{sd}=11.23$) from the crowdsourcing platform.
% \wys{
% Among them, 30 participants have a visualization background of below passing knowledge,
% while others' visualization literacy are above knowledgeable.}
% \yong{What is ``above knowledgeable''?}
We prescreened the candidates who are not from the United States to exclude the impact of cultural background and English proficiency.
% Our study was approved by our institution’s Institutional Review Board (\textit{IRB}). After the completion of the study and the review by us, e
Each participant was compensated with US \$3.28.
% Before the user study began, all participants will be redirected to the online study system designed by us. 
The study system was implemented via \textit{React.js} and deployed on \textit{Ubuntu 18.04.6 LTS}. The participants were asked to use a desktop device ($1920 \times 1080$ resolution) to perform the study tasks, ensuring the same quality representation of the target visual designs. 
% We recorded the participant's Prolific ID and provided them with the completion code at the very last of the study to trace every participant and improve the quality of the answers.

\subsection{Dataset}

For the quantitative study, we conducted the tasks based on the datasets generated illustrated in Sec. \ref{sec:metric_evaluation}.
Specifically, we prepared 36 datasets (small scale: 12, medium scale: 12, large scale: 12) for both visual designs. The datasets were randomly selected from the 405 generated datasets of each scale level. The target pair of line segments were randomly selected from the respective datasets.
Thus, each participant needed to perform 72 tasks (our approach: 36, baseline: 36) to evaluate the effectiveness of our approach for difference comparison compared to the baseline approach.

\subsection{Procedures}

% We conduct quantitative studies and a qualitative questionnaire to systematically evaluate \toolName. The respective procedures are as follows:

% \textbf{Quantitative study.}
The whole study of this part lasts about 20 minutes. In the beginning, we briefly introduced the purpose of the user study and what data will be collected during the procedure. After the participants consented to participate in the study and be redirected to the study system, we introduced the visual design of our approach (\textit{i.e.}, \toolName) and the baseline approach (\textit{i.e.}, slope graph), along with a tutorial about how to visually compare the state changes of two target data items using the two approaches. 
After the tutorial session, we conducted two mini-tests for each visual design, asking participants to choose the data item with larger differences from the two data items.
After passing the mini-tests, the formal quantitative study will begin. We collected the choice made by the participants and the time for each task. 
Specifically, we recorded their time used as the time between rendering a new chart and submitting the choice.
The target line segments to be identified were indicated by the arrows with different colors. 
% To avoid the participants comparing the line slopes via the arrows' vertical positions, we adjust the arrow directions the same as the target line segments to mitigate the effects.
% The interface of the online study system is provided in Appendix XX.

Note that the time of the tasks for \toolName\ included the time for the user to interact with the inner axis to assist them to identify the difference relationships, while the slope graph did not include any interaction time. 
Also, we designed the study as within-subject experiments to minimize the random noise from the participants. 
Meanwhile, we set a strict counter-balance order across all participants to mitigate the learning effects.
% To remind each participant to perform the study carefully, we terminate the test if the participant failed one of the mini-tests at the beginning of the study.
% In addition, we insert a warning page between the mini-test and the formal 36 studies to inform the participants the study and timer will begin and ask them to finish each task as quickly as possible.

% \textbf{Qualitative questionnaire.}
To collect the qualitative feedback of \toolName, we also conducted a post-study questionnaire.
The five questions are listed in Table \ref{table:2}.
We set the questions from three perspectives, \textit{i.e.}, effectiveness (Q1, Q2), usability (Q3), and visual design (Q4, Q5).
Aside from the open questions, the participants are also asked to provide their basic demographic information (\textit{i.e.}, age and gender) as well as their visualization literacy.
% which can better align with their performance during the quantitative experiments. Each participant will get a completion code after they filled out every open question in the questionnaire.

\begin{table}[tb]
\caption{The open questions in the questionnaire to collect the qualitative suggestions of \toolName.  Q1 and Q2 are used to evaluate the \toolName's \textit{effectiveness}; Q3 is for the \textit{usability} assessment; and Q4 and Q5 are used to evaluate the \textit{visual design}.} 
\centering
\begin{tabular}{c|p{0.8\columnwidth}}
\hline
Q1 & Do you think it is effective to compare the data items’ differences via Intercept Graph? Please explain why. \\ \cline{2-2} 
Q2 & Is Intercept Graph more suitable for comparing larger differences or small differences? Please explain why.                         \\ \cline{2-2} 
Q3 & Do you think it is easy and smooth to adjust the radius of the inner axis of Intercept Graph? Please provide more details. \\ \cline{2-2} 
Q4 & Do you think adjusting the inner axis radius is helpful for you to filter state changes of interest? Please explain why. \\ \cline{2-2} 
Q5 & Compared with slope graphs (\textit{i.e.}, using line slopes), do you think if it is easier/effective to compare state changes using Intercept Graph (\textit{i.e.}, line segment length)? Please provide more details on the pros and cons of both visual designs. \\ \hline
\end{tabular}
\label{table:2}
\end{table}

\begin{figure}[t]
\centering
\includegraphics[width=\columnwidth]{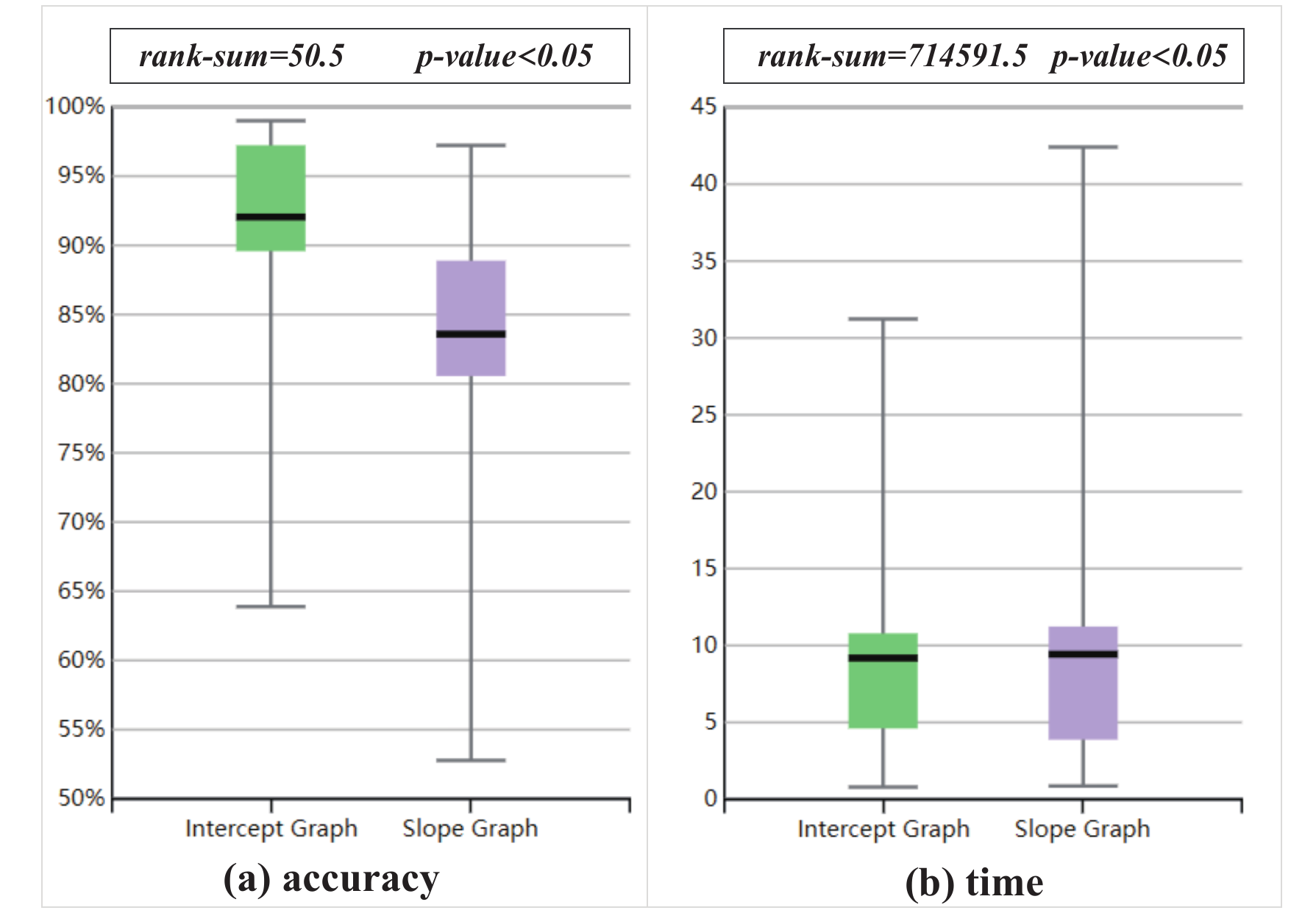}
\caption{
Metric evaluation to evaluate the accuracy and time cost of tasks using \toolName\ in comparison with the baseline approach slope graph. 
The mean of accuracy using \toolName\ significantly outperforms that of the slope graph, while the time costs for the two approaches are about the same.
The error bars are  95\% confidence intervals. Wilcoxon test statistics and corresponding p-values are reported at the top of each figure, which is much less than 0.05 for both metrics.
}
\label{fig:6}
\end{figure}

\subsection{Results}

We conduct the post-hoc analysis of the data and report the feedback of post-study questionnaire in this section.
% answers we collected from the user study. We report the results in this subsection.

% \textbf{Results for the quantitative study.}
\subsubsection{Accuracy and Time Cost}
We ran the non-parametric hypothesis test. \textit{i.e.}, the Wilcoxon test, for the post-hoc analysis, since the Shapiro-Wilks test shows that the data of accuracy and time cost is not normally distributed.
% its hypothesis of normality. 
Fig. \ref{fig:6} shows the results of statistical tests for time and accuracy: 
1) For the accuracy, \toolName\ significantly outperforms the baseline approach slope graphs (our approach: \textbf{92.05\%}, baseline approach: \textbf{83.58\%}). The \textit{p-value} indicates the significant difference of the performance between the two approaches. 
2) For the time cost, our approach \toolName\ (\textbf{9.18} seconds) is slightly better than slope graph (\textbf{9.43} seconds). The \textit{p-value} is still less than 0.05, indicating a significant difference in the time cost. 
% According to our observation, the reason is that the total time for \toolName\ included the time that participants interacted with the inner axis for more accurate and confident identification, while the time for slope graphs did not.

% \textbf{Qualitative feedback.}

\subsubsection{Feedback}
We collected the answers to the five open questions in the questionnaire. The answers are summarized as follows:
% \shaolun{need to be revised}

\textbf{Effectiveness.}
Most participants agreed that \toolName\ is more effective to compare the data items' differences than slope graphs (yes: 39, no: 9, not sure: 2). P15 commented that \toolName\ is more helpful to isolate the different lengths of line segments to compare their differences more easily. P32 mentioned, ``It has the effect of making it easier for me to see the difference value based on the line length directly. I have to compare the slopes of the lines when I'm using a slope graph, which is not intuitive.'' 
% \textbf{P23} also commented, ``At first it seems complicated over slope graph but it is more effective to compare data items when I get used to it.''
Also, most participants confirmed that \toolName\ is more suitable for comparing larger differences (large difference: 38, small difference: 8, not sure: 4). P21 commented that \toolName\ is better for larger differences because those line segments are much clearer around the circle center. Comparing small differences is challenging because the viewer will have a tough time deciding which line length is to be compared. P27 said, ``Large differences are immediately noticeable, whereas small differences are sometimes very unclear.''

\textbf{Usability.}    
Most participants agreed that it is easy and smooth to adjust the radius of the inner axis of \toolName\ (yes: 46, no: 3, not sure: 1). P19 praised the visualization, ``Absolutely, adjusting the inner axis was a lot smoother than I thought it would be. The system works very well.'' Also, P38 commented that the interaction is pretty smooth, and it allowed him to easily identify the larger difference.''

\textbf{Visual Design.}    
Most participants thought that adjusting the inner axis radius is helpful for them to filter state changes of interest (yes: 39, no: 5, not sure: 6). Meanwhile, P34 said, ``The interaction of the inner axis helps me in highlighting the data which have similar differences, and so adjusting it helps me to distinguish them and get a more accurate picture, which would become a problem in slope graphs.'' P4 also mentioned that it helped her to tell apart two line segments with similar lengths more easily than a slope graph.
Furthermore, most participants agreed that using line segment length (\textit{i.e.}, with \toolName) is easier and more effective to compare state changes than using the line slopes (\textit{i.e.}, with slope graph) (yes: 32, no: 8, equally good: 6, not sure: 4). P40 mentioned, ``I more like the Intercept Graph because of its design and practicality, as i think it was easier to tell how long the line was, however, the line slopes seemed a little harder to find what line is going to compare.'' 
% Also, P29 commented that he preferred intercept graph because he can easily tell the similar lines by dragging the inner axis. And the lines in the intercept graph look more clear and easy to tell than in the slope graph when there are a bunch of lines.

\textbf{Suggestions.} 
Despite the positive feedback, several participants also provided suggestions on further improving \toolName.
P16 commented that \toolName\ may be able to mitigate the visual clutter in parallel coordinates, making users able to recognize the data distributions more easily.
% P39 also suggested that \toolName\ could support the interaction on mobile phones, instead of desktop-based interaction only.
P43 described her experience of exploring \toolName\ via mobile phones. She then suggested that \toolName\ may support the automatic determination of the radius of the inner axis, especially when the users feel inconvenient to drag the inner axis.
% \yong{Pls check it.}

% \yong{pls check my Chinese comments!}

\subsection{Summary}

Through the user study, we conclude that \toolName\ can effectively assist participants in identifying the relationships of a pair of state changes, making it more accurate than the baseline approach slope graph (our approach: \textbf{92.05\%}, baseline approach: \textbf{83.58\%}) based on a similar time costs (our approach: \textbf{9.18} seconds, baseline approach: \textbf{9.43} seconds). Meanwhile, through the questionnaire, most participants agreed that our approach \toolName\ can better support the filtering and comparison of the line segments based on the interaction with the inner axis.
% , especially when the number of line segments becomes large or the differences in data items are too close to tell.

\section{discussion}

In this section, we discuss the lesson we learned and its limitations.
% , and generalization.

\textbf{Lessons Learnt.}
We learned a valuable lesson from the development and implementation of \toolName{}.
We found that it is important to make the visualization design and interactions straightforward.
% through our iterative project meetings, 
For our initial design, we use extra visual components to interact with \toolName{}, \textit{i.e.}, an input box for filtering and a slider bar for comparison.
% which however
% ) to support the user interaction.
But our user feedback shows that it can confuse users when interacting with \toolName{}.
% makes the visualization too complex to understand . 
% To simplify \toolName, 
Thus,
we
% integrated all the interaction components into 
further allow users to interact with \toolName{} by directly dragging
the inner axis, making the filtering and comparison of data items more intuitive and time-saving.
% , which can enable users to drag directly. 
% We found this modification can make the comparison more intuitive and time-saving.
% Second, through our literature review of the papers for state change comparison, we realized that researchers show more interest in observing and studying those data items with relatively large state changes. 
% In those research papers, these data items usually have the most significant effects or are anomalous, which are worthy of being studied further. 
% \yong{Add another lesson.}

\textbf{Limitations.}
(1) \textit{Visual scalability: }
According to the feedback we collected in the user study, participants reported that \toolName\ is effective in facilitating the filtering and comparison state changes when the number of data items is small-scale (\textit{i.e.}, 5$\sim$25) or medium-scale (\textit{i.e.}, 100$\sim$120). However, the above two features can be affected when the number of data items is large-scale (\textit{i.e.}, 200$\sim$220), as the data items with relatively small state changes will cause visual clutter on the circumference of the outer axis.
Thus, \toolName\ is more appropriate for comparing large state changes.
2) \textit{Usage on mobile devices:}
\toolName\ can support a smooth filtering and accurate comparison by leveraging the mouse to drag the inner axis.
% a flexible mouse interaction (\textit{i.e.}, 
% dragging the inner axis
% % ).
However, the interaction will be hindered if mouse-dragging is not allowed in some scenarios.
For example, for mobile phone users, it is not convenient to drag the inner axis,
since touching cannot enable an accurate selection and dragging of visual elements
% by users may trigger other  (e.g., the display of the data item's name label) 
due to the limited screen space.
% making the interaction difficult and inflexible.

% \textbf{Generalization.}
% Our usage scenarios show how \toolName\ can enable effective state change comparison across different timestamps.
% But its usage is not limited to such temporal state change comparison and can be extended to compare any data differences, for example, comparing the customers' rating difference between two different restaurants.

% The usage scenario demonstrates the usefulness of \toolName\ for state change comparison. But its applications are not limited to the state change comparison and can also be extended to other application scenarios that need a comparison between two data series,
% instead of the applications for temporal changes (e.g., the initial and final values) only.
% For example, \toolName\ can also support the comparison of the customers' ratings between two different restaurants. 
\section{Conclusion}

We present \toolName, a novel visual design to facilitate the comparison of state changes. 
We encode the state difference magnitude by the length of each line segment.
Building upon the mathematical theorem, through the interaction with the inner axis, \toolName\ can intrinsically support smooth filtering of the state difference.
Meanwhile, the inner axis allows a magnification of the difference of state changes, making the comparison more flexible and accurate.
We present two usage scenarios to demonstrate the usefulness of \toolName\  for different applications.
Also, the metric evaluation proved that \toolName\ can effectively mitigate the line crossings and enhance the difference perception over the baseline approach, \textit{i.e.}, slope graph.
Furthermore, we conducted a user study with 50 participants to evaluate the performance of humans using \toolName\ in comparison with the slope graph. The results confirmed the usefulness and effectiveness of our approach in state change comparison.
% \yong{This paragraph talks more about the evaluation. But you also need to further highlight the approach itself, e.g., its visual encoding, its major advantages, which should be extended further here!}

In future work, we plan to investigate how the visual scalability of \toolName{} can be improved to mitigate the visual clutter when visualizing an extremely large number of data items.
% , e.g., through edge bundling. 
Also, it would be interesting to explore how to \textit{automatically} determine the optimal inner circular radius for different datasets and facilitate a more efficient comparison of state changes.

\ifCLASSOPTIONcompsoc
  \section*{Acknowledgments}
\else
  \section*{Acknowledgment}
\fi

This research was supported by the Singapore Ministry of Education (MOE) Academic Research Fund (AcRF) Tier 1 grant (Grant number: 20-C220-SMU-011).
% The authors would like to thank Songheng Zhang, Xiaolin Wen, Yanna Lin, and Cheng Chen for
% % constructive criticism of the manuscript.
% proofreading the paper.

\ifCLASSOPTIONcaptionsoff
  \newpage
\fi



\bibliographystyle{IEEEtran}
\bibliography{main.bib}


% \end{thebibliography}


\begin{IEEEbiography}[{\includegraphics[width=1in,height=1.5in,clip,keepaspectratio]{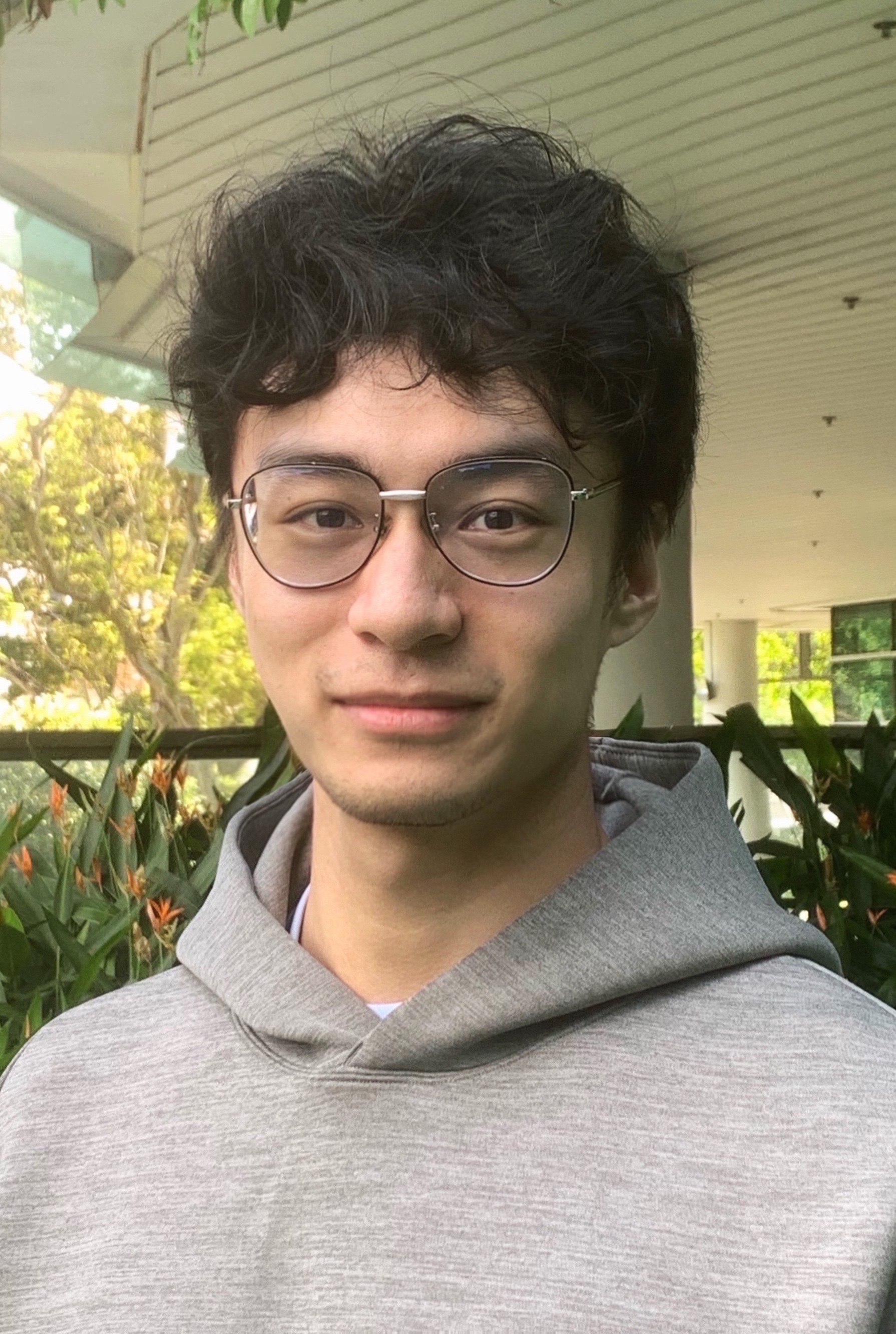}}]{Shaolun Ruan} is currently an Ph.D. student in School of Computing and Information Systems at Singapore Management University. His major research interests include data visualization and human-computer interaction. He received his bachelor degree from University of Electronic Science and Technology of China majoring in Information Security at School of Computer Science and Engineering in 2019. For more details, please refer to \url{https://shaolun-ruan.com/}.
\end{IEEEbiography}

\begin{IEEEbiography}[{\includegraphics[width=1in,height=1.5in,clip,keepaspectratio]{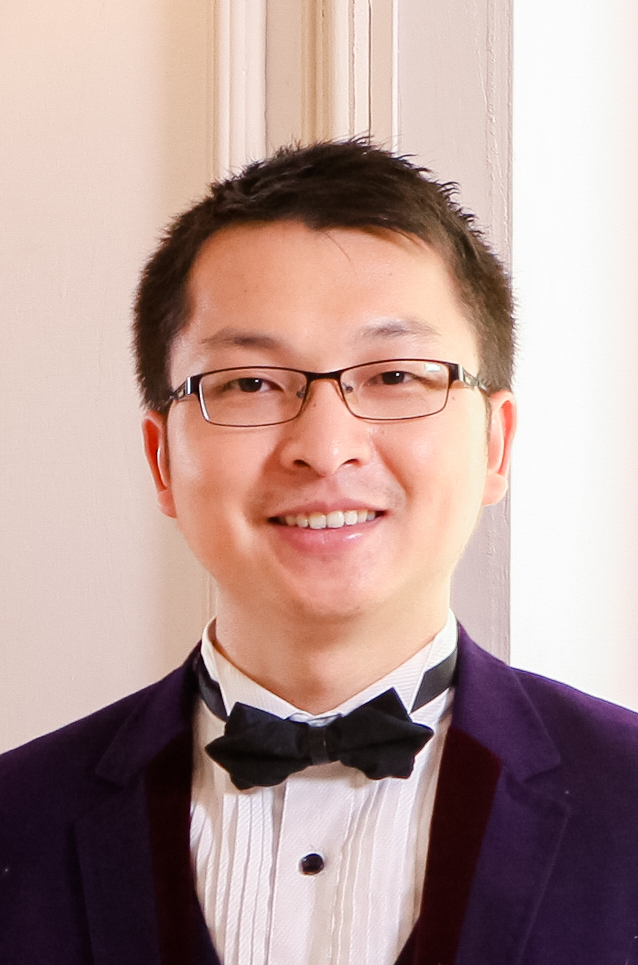}}]{Yong Wang} is currently an assistant professor in School of Computing and Information Systems at Singapore Management University. His research interests include data visualization, visual analytics and explainable machine learning.
He obtained his Ph.D. in Computer Science from Hong Kong University of Science and Technology in 2018. He received his B.E. and M.E. from Harbin Institute of Technology and Huazhong University of Science and Technology, respectively. For more details, please refer to \url{http://yong-wang.org}.
\end{IEEEbiography}

\begin{IEEEbiography}[{\includegraphics[width=1in,height=1.5in,clip,keepaspectratio]{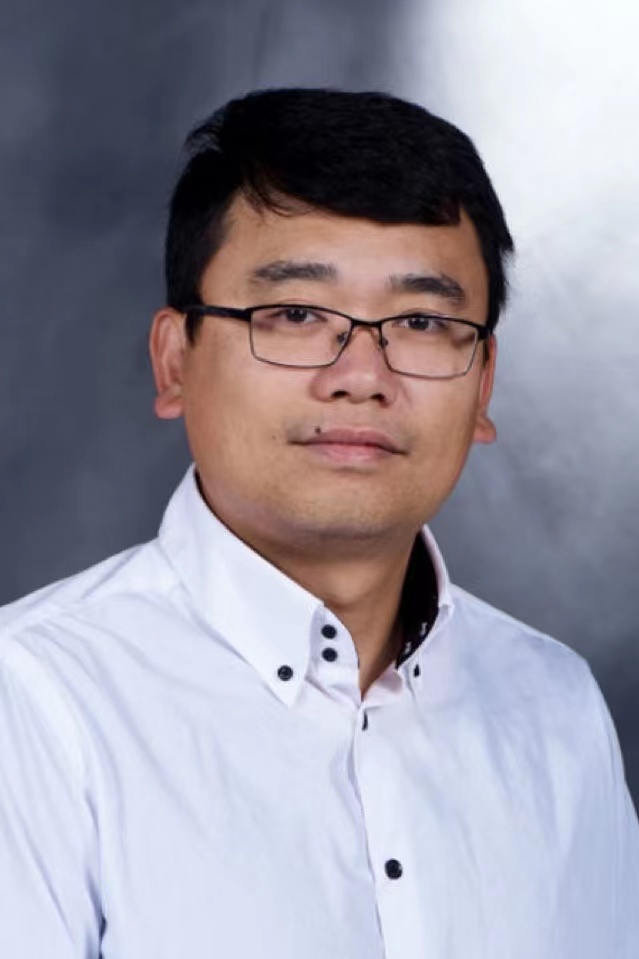}}]%
{Qiang Guan}
Dr. Qiang Guan is an assistant professor in Department of Computer Science at Kent State University, Kent, Ohio. Dr. Guan is the direct of Green Ubiquitous Autonomous Networking System lab (GUANS). He is also a member of Brain Health Research Institute (BHRI) at Kent State University. He was a computer scientist in Data Science at Scale team at Los Alamos National Laboratory before joining KSU. His current research interests include fault tolerance design for HPC applications; HPC-Cloud hybrid system; virtual reality; quantum computing systems and applications. For more details, please refer to \url{http://www.guans.cs.kent.edu/}
\end{IEEEbiography}

% \clearpage

% \appendix
% \input{source/10_appendix}

%\newpage

%\begin{IEEEbiographynophoto}{Qiang Guan}


%\vfill


\end{document}



% TODO: 
% 给case study的内容加R1等requirement编号
% 1. Temporal View -> \textit{Temporal View}
% 2.IBM Q -> \textit{IBM Q}
% 3. System 部分删去了一些交互模块的annotation, 有teaser之后写进去
% 4. ibmq_lima -> \textit{ibmq_lima}
% 5

% \input{source/1-intro}
% \input{source/2-relatedwork}
% \input{source/2.1-background}
% \input{source/2.2-informing_the_design}
% \input{source/3-system}
% \input{source/4.1-case_study}
% \input{source/4.2-user_interview}
% \input{source/5-discussion}
% \input{source/6-conclusion}

% The authors wish to thank A, B, and C. This work was supported in part by
% a grant from XYZ.}

%\bibliographystyle{abbrv}
\bibliographystyle{abbrv-doi}
%\bibliographystyle{abbrv-doi-narrow}
%\bibliographystyle{abbrv-doi-hyperref}
%\bibliographystyle{abbrv-doi-hyperref-narrow}
% \Urlmuskip=0mu plus 1mu
% \bibliography{template}

% \newpage
% \clearpage

\appendix

% \appendices
\section{Details of the User Study}
\label{sec:appendix_1}

We proposed an online web-based system for the user study.
Our study was approved by our institution’s Institutional Review Board (\textit{IRB}). 
The pipeline of the user study is shown in Figure \ref{fig:appendix_1}.
Before the user study began, all participants will be redirected to the online study system designed by us. 
In the very beginning, we recorded the participant's Prolific ID and provided them with the completion code at the very last of the study to trace every participant and improve the quality of the answers. The participants are required to complete all mini-tests to improve the experiment quality.
The 72 tasks and the questionnaire are the subsequent procedures.
Note that all participants are divided into two groups for a strict counter-balance order to mitigate the learning effects, as shown in the two paths (i.e., Task Sequence 1 and Task Sequence 2) in Figure \ref{fig:appendix_1}.
All choices and questions are required for the completion.
Each participant will get a completion code after they answer all open questions in the questionnaire.

Note that to avoid the participants comparing the line slopes via the arrows' vertical positions, we adjust the arrow directions the same as the target line segments to mitigate the effects.
To remind each participant to perform the study carefully, we terminate the test if the participant failed one of the mini-tests at the beginning of the study (Figure \ref{fig:appendix_1}(B)).
In addition, we insert a warning page between the mini-test and the formal 36 studies to inform the participants the study and timer will begin and ask them to finish each task as quickly as possible (Figure \ref{fig:appendix_1}(A)).

\begin{figure*}[b]
\centering
\includegraphics[width=0.9\linewidth]{appendix_1.pdf}
\caption{The pipeline of the system for user study, consisting of the screenshots of the system interfaces. Each participant is required to complete all tasks and the questionnaire to receive the completion code at the very last of the study. Each participant was compensated with US \$3.28.}
\label{fig:appendix_1}
\end{figure*}

% \begin{figure*}[b!]
% \centering
% \includegraphics[width=0.5\linewidth]{figures/appendix_1.pdf}
% \caption{
% The pipeline of the system for user study, consisting of the screenshots of the system interfaces. Each participants are required to complete all tasks and the questionnaire to receive the completion code at the very last of the study. Each participant was compensated with US \$3.28.}
% \label{fig:appendix_1}
% \end{figure*}

\section{Usage of Our Open-source Package for Intercept Graph}

To support the easy implementation of \toolName, we develop a JavaScript function and make it open-public via the URL: \textcolor{blue}{\url{https://www.npmjs.com/package/interceptgraph}}. The open-sourced package supports the accurate and smooth comparison of state changes for the general public.

We provide a useful API for the quick implementation of \toolName.
Specifically, we define two parameters to build the chart:

\begin{itemize}
    \item \textbf{\textit{svg\_id:}} the identification of the SVG container to draw \toolName. The parameter only accepts the \textit{id} of the DOM element. Note that the width and height of the SVG should be specified within the tag using \textit{width} and \textit{height} parameters.
    
    \item \textbf{\textit{src:}} the data source to be rendered. The parameter only accept \textit{CSV} data file as input. Using the path name to the \textit{CSV} file for the parameter. Note that the \textit{CSV} data file should be formatted as follows: 1) no headers for the dataset; 2) the order of the data field: \textit{name}, \textit{data series 1}, \textit{data series 2}; 3) use ``,'' for separating the attributes.
\end{itemize}

% For \textit{ES 2015}, the user should install the package and then import the module using `import * from 'interceptgraph'`. Call the function using `interceptgraph\_build()` to draw the chart.
% For \textit{Common.js}, the user should import the module using the HTML tag. Call the method of object \textit{ig} (e.g., `ig.interceptgraph\_build()`) to draw the chart. 
An example of the usage of the package is shown in Figure \ref{fig:appendix_3}.

\begin{figure}[hbtp]
\centering
\includegraphics[width=\columnwidth]{appendix_3.pdf}
\caption{
An example is to call the rendering function to draw an Intercept Graph conveniently and quickly. The user needs to mount the node of the SVG element first, and then render the chart into the SVG element. The size of the SVG element is required to be specified in advance instead of including it in a style sheet.}
\label{fig:appendix_3}
\end{figure}

\section{Generated Datasets for Metric Evaluation}
In Section 6, we use randomly-generated datasets to evaluate the metrics, i.e., line crossings and intensity ratio. For the dataset generation, we define three scale levels (i.e., small scale, medium scale, and large scale) and 81 normal distributions to perform the metric evaluation. In this section, we showcase some of the charts (i.e., our approach: \toolName, baseline approach: slope graph) driven by the datasets we generated in advance, as shown in Figure \ref{fig:appendix_2}. 

The procedure for dataset generation is as follows: first, we defined three scale levels (i.e., the small-scale, the medium-scale, and the large-scale) of the number of data items. For each scale level, an increment of 5 data items was set to cover a wider range.
Assume that the axis ranges of the two data series were from 0 to 100. For each \textit{randomly-generated} dataset, we tried to generate all possible distributions for the given number of data. We applied the Gaussian Distribution to the data points generation and set three types of means and three types of standard deviations to the data series (i.e., $d_1$ and $d_2$), respectively.
We, therefore, generated 1215 randomly-generated datasets in total: 3 scale levels with 5 scale numbers per level $\times$ 9 Gaussian Distributions of data series $d_1$ $\times$ 9 Gaussian Distributions of data series $d_2$.

We evaluate \toolName\ in comparison with the slope graph that \toolName\ can effectively mitigate the line crossings compared to the slope graph, especially for the datasets with a large number of data items (e.g., over a medium scale of 200). Meanwhile, \toolName\ can magnify the ratio of the length of two state changes, making it more intuitive to tell apart the relationship between two state changes.

\begin{figure*}[hbtp]
\centering
\includegraphics[width=0.95\linewidth]{appendix_2.pdf}
\caption{
The comparison of \toolName\ and slope graph. We use the same data to evaluate \toolName\ and the slope graph in each experiment. The Gaussian Distribution shown in the chart is a part of the whole set (i.e., 12 out of 81). $\mu_{1}, \sigma_{1}$ are for Gaussian Distribution of data series 1, while $\mu_{2}, \sigma_{2}$ are for Gaussian Distribution of data series 2. The other distribution parameters in each row are set as the first value in each parameter set, as shown in Table 1.
}
\label{fig:appendix_2}
\end{figure*}